\documentclass[5p, onecolumn]{elsarticle}

\journal{Journal of Systems and Software}
\usepackage[dvipsnames]{xcolor}
\usepackage{natbib}
\usepackage{amsmath,amssymb,amsfonts}
\usepackage{algorithmic}
\usepackage{url}
\usepackage{graphicx}
\usepackage{array}
\usepackage{multirow}
\usepackage{verbatim}
\usepackage{lipsum}
\usepackage{hyphenat}
\usepackage{float}
\usepackage{todonotes}
\usepackage{tcolorbox}
\usepackage{enumitem}
\usepackage{balance}
\usepackage{todonotes}
\usepackage{threeparttable}
\usepackage{listings}
\newcolumntype{?}{!{\vrule width 1pt}}
\usepackage{fourier} 
\usepackage{caption}
\usepackage{subcaption}
\usepackage{booktabs}
\usepackage{hyperref}
\usepackage{footmisc}
\usepackage[export]{adjustbox}
\usepackage{url}
\usepackage{arydshln}
\usepackage{pifont}
\usepackage{tikz}
\usepackage{tablefootnote}

\usepackage{hyperref} 
\hypersetup{hidelinks} 

\usepackage{soul}
\setuldepth{Berlin}

\newcommand{\toolapp}{AppDynamics}
\newcommand{\tooldog}{Datadog}
\newcommand{\tooldyn}{Dynatrace}
\newcommand{\toolhay}{Haystack}
\newcommand{\tooljae}{Jaeger}
\newcommand{\toolnew}{New Relic}
\newcommand{\toolope}{OpenTelemetry}
\newcommand{\toolsen}{Sentry}
\newcommand{\toolspl}{Splunk}
\newcommand{\toolzip}{Zipkin}
\newcommand{\toolela}{ElasticAPM}
\newcommand{\tooloce}{Ocelot}
\newcommand{\toolint}{Instana}
\newcommand{\toollig}{LightStep}
\newcommand{\toolsky}{SkyWalking}
\newcommand{\toolsta}{StageMonitor}
\newcommand{\toolvmw}{Wavefront}

\definecolor{darkmagenta}{rgb}{0.55, 0.0, 0.55}
\definecolor{darkgreen}{RGB}{6, 46, 3}
\definecolor{amber}{rgb}{1.0, 0.75, 0.0}
\definecolor{ao(english)}{rgb}{0.0, 0.5, 0.0}

\usepackage{color}
\newcommand\ra[1]{\textcolor{black}{#1}}
\newcommand\rb[1]{\textcolor{black}{#1}}

\newcommand\rsa[1]{\textcolor{black}{#1}}
\newcommand\rsb[1]{\textcolor{black}{#1}}

\newcommand\issuetopic[1]{\textcolor{red}{#1}}
\newcommand\benefittopic[1]{\textcolor{ao(english)}{#1}}

\newcolumntype{L}[1]{>{\raggedright\let\newline\\\arraybackslash\hspace{0pt}}p{#1}}
\newcolumntype{C}[1]{>{\centering\let\newline\\\arraybackslash\hspace{0pt}}p{#1}}
\newcolumntype{R}[1]{>{\raggedleft\let\newline\\\arraybackslash\hspace{0pt}}p{#1}}
\newcommand*\rot{\rotatebox{90}}
\newcommand*\rotback{\rotatebox{270}}

\newcommand{\issue}{{\color{black}\textbf{\normalsize--}}}
\newcommand{\benefit}{{\color{black}\textbf{\normalsize+}}}
\newcommand{\implied}{\color{black}\ding{53}}
\newcommand{\yes}[1][]{\ding{53}\textsuperscript{#1}}

\def\BibTeX{{\rm B\kern-.05em{\sc i\kern-.025em b}\kern-.08em
    T\kern-.1667em\lower.7ex\hbox{E}\kern-.125emX}}

\mathchardef\UrlBreakPenalty=1000
\mathchardef\UrlBigBreakPenalty=1000

\begin{document}
\begin{frontmatter}


\title{Open Tracing Tools: Overview and Critical Comparison}

\author[FHV,BZ]{Andrea Janes}
\ead{andrea.janes@fhv.at}
\author[TUNI,OULU]{Xiaozhou Li}
\ead{xiaozhou.li@oulu.fi}
\author[OULU]{Valentina Lenarduzzi}
\ead{valentina.lenarduzzi@oulu.fi}

\address[FHV]{FHV Vorarlberg University of Applied Sciences, Austria}
\address[BZ]{Free University of Bozen-Bolzano, Italy}
\address[TUNI]{Tampere University, Finland}
\address[OULU]{University of Oulu, Finland}

\begin{abstract}
\textit{Background.} Coping with the rapid growing complexity in contemporary software architecture, tracing has become an increasingly critical practice and been adopted widely by software engineers. By adopting tracing tools, practitioners are able to monitor, debug, and optimize distributed software architectures easily. However, with excessive number of valid candidates, researchers and practitioners have a hard time finding and selecting the suitable tracing tools by systematically considering their features and advantages. \\
\textit{Objective.} To such a purpose, this paper aims to provide an overview of popular Open  tracing tools via comparison. \\ 
\textit{Method.} Herein, we first identified \ra{30} tools in an objective, systematic, and reproducible manner adopting the Systematic Multivocal Literature Review protocol. Then, we characterized each tool looking at the 1) measured features, 2) popularity both in peer-reviewed literature and online media, and 3) benefits and issues. We used topic modeling and sentiment analysis to extract and summarize the benefits and issues. Specially, we adopted ChatGPT to support the topic interpretation. \\
\textit{Results.} As a result, this paper presents a systematic comparison amongst the selected tracing tools in terms of their features, popularity, benefits and issues. \\
\textit{Conclusion.} The result mainly shows that each tracing tool provides a unique combination of features with also different pros and cons. The contribution of this paper is to provide the practitioners better understanding of the tracing tools facilitating their adoption.
\end{abstract}


\end{frontmatter}

\section{Introduction}
\label{sec:intro}
\noindent In \textit{software} engineering, the outcome of the engineering process is invisible \cite{Brooks1987,Fenton1998}. As a consequence, it is difficult to understand progress and to reason about the produced output \cite{BCS2004}. This is particularly complicated when developing systems that consist of many components. Today's trend of developing systems interacting with components deployed in the cloud or based on microservice architectures only exacerbates this problem.

One way to cope with invisibility is through measurement. Measurement enhances observability as it provides the data needed to understand the internal states of systems and its components \cite{Kalman1960}. Measurement is defined as the process of assigning numbers or symbols to the attributes of real-world entities to describe them using clearly defined rules \cite{Finkelstein1984,Fenton1998}.

\begin{sloppypar}
Medicine distinguishes between the measuring instruments that need to \textit{look inside} a patient and those that do not. For example, blood pressure can be measured directly via an arterial catheter (called \textit{invasive}) or by placing a stethoscope on an artery, pumping up a cuff placed around the arm, and reading blood pressure on a special meter called a sphygmomanometer (such an approach is called \textit{non-invasive}). Following the same terminology, measurement of software can be invasive or non-invasive: we can distinguish methods that require to modify the source code of the measured system (e.g., logging relevant events) or methods that consider the observed system a black box and measure how it interacts with the environment.
\end{sloppypar}

The two terms often used in software measurement are \textit{tracing} and \textit{telemetry}. Tracing, as the word \textit{trace}, means ``a mark or line left by something that has passed'' \cite{Trace}, is often used by developers to log what has happened and to understand if software is working as expected or not. While tracing is also measurement, the term emphasizes that relevant events are logged---together with the time of occurrence---during the execution of software. In contrast, counting the lines of code of a class is not tracing, it is only measurement.

The second frequent term is telemetry: the word is composed by the Greek adjective \textit{tele} (remote) and the word \textit{me\-tron} (measure) and means to measure something and transmitting the results to a distant station \cite{Telemeter}. Such an approach is often needed when a large quantity of data is collected and it cannot be processed in situ; or, if data is collected from several sources (as in a distributed system) and it is necessary to collect the data in one place to obtain the complete picture of what is happening in the system. Another term often used in a distributed context is \textit{distributed tracing}, where a \textit{trace} represents the ``whole journey of a request as it moves through all of the services of a distributed system'' \cite{distributed-tracing} and the term \textit{span} describes the part of the trace belonging to one service: a span represents a logical unit of work in completing a user request within \textit{one} service; combining all spans describes one request across all services. Distributed tracing, implicitly, includes telemetry, since the data collected from various points needs to be transmitted for processing to another device.

Tracing is used in a variety of cases, e.g., to locate the cause why a system does not meet performance requirements or where failures occur. It is part of the toolkit used by software engineers to monitor, debug, and optimize distributed software architectures, such as microservices or serverless functions. Researchers and practitioners need to be aware of the tools currently in use and what features they possess. Tool vendors and (potential) tool producers need to understand how popular and adopted their tools are, as well as for through features they distinguish themselves from the competitors. 

\begin{sloppypar}
For this purpose, this paper aims to obtain an overview of tracing tools and to perform a critical comparison among them, focusing on those that are used among researchers and practitioners and are available on the market using an Open Source license. To achieve this objective, we first identified \ra{30} tools in an objective, systematic, and reproducible manner adopting a Systematic Multivocal Literature Review approach \cite{Garousi18a}. Then, we characterized each tool looking at:
\end{sloppypar}
\begin{itemize}
    \item What the tracing tool is able to measure (distinctive features);
    \item Popularity (both in peer-reviewed literature and online media);
    \item Advantages (benefits) disadvantages (issues) reported by researchers and practitioners in social media articles using 
   \begin{enumerate}
       \item topic modeling and sentiment analysis techniques for the benefits and issues extraction,
       \item ChatGPT~\cite{chatgpt} to support the topic interpretation.
   \end{enumerate}
\end{itemize}
 
The key results of the study show that, among the considered tools, \ra{the features provided vary.}

\ra{Therein, 10 of them are considered popular based on the volume of social media discussion. These are the ones we took into account for the benefits and issues analysis due to the lack of data from the others. There are six main criteria on which practitioners mainly have opinions. The results show that only a very limited number of tools provide all the considered features but none of the tools are perceived positively or negatively in all the aspects. }


\ra{To sum up, the main contribution of our work is represented by the large analysis up to date on the overview and comparison of the capabilities of Open Tracing Tools that implement Open Tracing API. Specifically, we advanced the current state of the art in four different manners:}

\begin{enumerate}
    \item \ra{By providing results from a objective, systematic, and reproducible approach and a detailed replication package\footref{Package} with the data and scripts used to conduct our study and that can be used by the research community and practitioners to replicate and build upon it;}
    
    \item \ra{By providing a comparison of the distinctive features each tool possessed, which may be used by practitioners as a way to select the most suitable tool(s) based on the specific project needs;}
    
    \item \ra{By providing an overview  about how much each tool is known for the researchers and practitioners considering how long have they been on the market; }
    
    \item \ra{By investigating the benefits and issues among the considered tools, which can inform tool vendors about the limitations of the current solutions available the market, other than making practitioners aware of how to benefit more from the combined capabilities of the considered tools;}
    
\end{enumerate}

The remainder of this paper is structured as follows. Section~\ref{sec:ML} presents the process we followed to identify the Open Tracing Tools studied in this paper, while Section~\ref{sec:EmpiricalStudy} describes the empirical study we conducted. Section~\ref{sec:Results} describes the obtained results, while \ra{Section}~\ref{sec:Discussion} discusses them. Section~\ref{sec:TV} highlights the threats to validity of this work, \ra{Section}~\ref{sec:RW} the related work, and \ra{Section}~\ref{sec:Conclusion} draws conclusions and future works.

\section{Systematic Open Tracing Tools Selection}
\label{sec:ML}
\noindent To identify a list of Open Tracing Tools in an objective, systematic, and reproducible manner, we adopted the approach of a Systematic Multivocal Literature Review (MLR)~\cite{Garousi18a}. The MLR process~\cite{Garousi18a} includes both peer reviewed as well as gray literature and the different perspectives between practitioners and academic researchers are taken account in the results. A MLR emphasizes the inclusion of gray literature in the data collection process for topics with a strong interest by practitioners. MLR classifies contributions as \textit{academic literature} in case of peer-reviewed papers and as \textit{gray literature} other types of content like blog posts, white-papers, pod casts, etc.

\begin{figure}[ht]
    \centering
    \includegraphics[width=.4\textwidth]{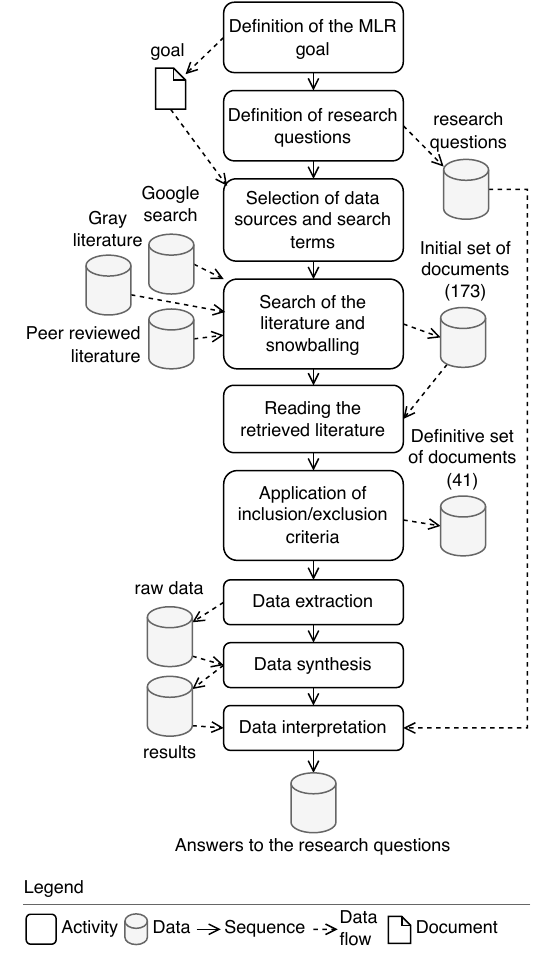}
    \caption{Overview of the followed MLR process (adapted from Fig. 7 in \cite{Garousi18a})}
    \label{fig:mlrprocess}
\end{figure}

\begin{sloppypar}
The process is divided into different phases with the main steps we followed depicted in Figure \ref{fig:mlrprocess}. We started with the definition of the overall goal of the study and the formulation of research questions. The goal and the research questions determine the selection of the data sources and the search terms. The literature review is executed searching the literature and performing snowballing \cite{Wohlin2014}. After reviewing the initial set of documents, applying the inclusion/exclusion criteria, and evaluating the quality and credibility of sources, we obtained 41 documents. Based on a defined data extraction scheme, we extracted data useful to answer the defined research questions, and---through data synthesis and inter\-pre\-tation---we obtained the answers to the research questions.
\end{sloppypar}

Formulated as a GQM measurement goal \cite{Basili2014}, the objective of this paper can be described as follows: ``Analyze the current literature about Open tracing tools for the purpose of characterization, with respect to its distinctive features, its popularity, and benefits and issues, from the point of view of a software developer in the context of a software development organization.''

Following the guidelines to formulate questions along the GQM (goal, question, metric) paradigm \cite{Basili1994gqm}, we devise the below research questions operationalizing ``utility'' into four dimensions: ability to measure, popularity, benefits, issues. Consequently, we formulated the following research questions:

\begin{description}[itemsep=-1ex]
    \item[RQ$_1$:] Which distinctive features do the tools possess?
   \item[RQ$_2$:] How popular are the identified tools?
    \item[RQ$_3$:] Which benefits do the identified tools claim to achieve?
    \item[RQ$_4$:] Which issues do the identified tools introduce?
\end{description}

With \textbf{RQ$_1$}, we aimed at characterizing each tool with respect to their distinctive features to provide a clear comparison of the differences between them. Though many tools provide similar functionalities, each one also provides unique features compared to the others.  Then, in \textbf{RQ$_2$}, we investigated the popularity of each tool in terms of adoption both in academia and among the developers communities considering when the firs version was released on the market.  Once characterized each tool, we proceeded with a finer-grained investigation considering benefits (\textbf{RQ$_3$}) and issues (\textbf{RQ$_4$}) to understand the unique advantages and drawbacks of each tool so that they can adopt the ones best suiting their needs.

\noindent To obtain a high recall and include as many papers as possible, we used the following broad search string to retrieve literature about open tracing tools:

\begin{center}
\textbf{(opentracing OR ``open tracing'') AND tool*}
\end{center}
We used the asterisk character (*) to capture possible term va\-riations such as plurals and verb conjugations. The search terms were applied to all the fields (i.e. title, abstract, and keywords), so as to include as many works as possible. 

\textbf{Peer-reviewed literature search}. We considered the papers indexed by several bibliographic sources, namely: 
ACM digital Library\footnote{\url{https://dl.acm.org}}, 
IEEEXplore Digital Library\footnote{\url{https://ieeexplore.ieee.org}},
Science Direct\footnote{\url{https://www.sciencedirect.com}}, Scopus\footnote{\url{https://www.scopus.com}}, 
Google Scholar\footnote{\url{https://scholar.google.com}}, 
CiteseerX\footnote{\url{https://citeseer.ist.psu.edu}}, Inspec\footnote{\url{https://iee.org/Publish/INSPEC/}}, and 
Springer link\footnote{\url{https://link.springer.com/}}.
The search was conducted in \ra{November 2022}
and all raw data are presented in the replication package (Section~\ref{subsec:Replicability}).

\textbf{Gray literature search}. We adopted the same search terms for retrieving gray literature from online sources as we did for peer-reviewed ones. We performed the search using four search engines: Google Search\footnote{\url{https://www.google.com/}}, Twitter\footnote{\url{https://twitter.com/}}, Reddit\footnote{\url{https://www.reddit.com/}} and Me\-dium\footnote{\url{https://medium.com}}. The search results consisted in books, blog posts, forums, websites, videos, white-paper, frameworks, and podcasts. This search was performed in \ra{November 2022}.

\textbf{Snowballing}. Snowballing refers to using the reference list of a paper or the citations to the paper to identify additional papers \cite{Wohlin2014}. We applied backward-snowballing to the academic literature to identify relevant papers from the references of the selected sources. Moreover, we applied backward-snowballing for the gray literature following outgoing links of each selected source. 

\textbf{Application of inclusion and exclusion criteria}. Based on guidelines for Systematic Literature Reviews \cite{Kitchenham2007}, we defined \textit{inclusion and exclusion} criteria. 
\rb{We included tools that implement Open Tracing APIs because of their importance for tool providers and in particular, for tools aiming at architectural reconstruction and observavility}.
We excluded tools that could not be downloaded or installed, tools with no documentation on how to install or deploy them, as well as tools without a web site.


\textbf{Evaluation of the quality and credibility of sources}. Differently than peer-reviewed literature, gray literature does not go through a formal review process, and therefore its quality is less controlled. To evaluate the credibility and quality of the selected gray literature sources and to decide whether to include a gray literature source or not, we extended and applied the quality criteria proposed by~\cite{Garousi18a} considering the authority of the producer, the applied methodology, objectivity, date, novelty, and impact. 

Two authors assessed each source using the aforementioned criteria, with a binary or three-point Likert scale, depending in the criteria itself. In case of disagreement, we discussed the evaluation with the third author that helped to provide the final assessment. 

Table \ref{tab:Context} lists the outcome of the systematic tool selection, i.e., the tools that we identified through the above described process.


\begin{table}[ht]
    \setlength{\tabcolsep}{4pt}
    \centering
    \footnotesize
    \caption{\ra{The 30 retrieved tools}}
    \label{tab:Context}
    \begin{tabular}{L{2cm}|l} \hline 
    \textbf{Tool name} &\textbf{Web site} \\ \hline 
\ra{Appdash}& https://github.com/sourcegraph/appdash\\\hdashline[1pt/1pt]
Appdynamics& https://www.appdynamics.com/\\\hdashline[1pt/1pt]
\ra{Containiq}& https://www.containiq.com/\\\hdashline[1pt/1pt]
\textsc{Datadog} & https://www.datadoghq.com/\\\hdashline[1pt/1pt]
\ra{Dynatrace}& https://www.dynatrace.com/\\\hdashline[1pt/1pt]
Elasticapm & https://www.elastic.co/\\\hdashline[1pt/1pt]
\ra{Grafana tempo} & https://grafana.com/oss/tempo/\\\hdashline[1pt/1pt]
\ra{Haystack}& https://expediadotcom.github.io/haystack/\\\hdashline[1pt/1pt]
\ra{Honeycomb.io}& https://www.honeycomb.io/\\\hdashline[1pt/1pt]
\ra{Hypertrace}& https://www.hypertrace.org/\\\hdashline[1pt/1pt]
Instana& https://www.instana.com/\\\hdashline[1pt/1pt]
Jaeger& https://www.jaegertracing.io/\\\hdashline[1pt/1pt]
\ra{Kamon}& https://kamon.io/\\\hdashline[1pt/1pt]
Lightstep& https://lightstep.com/\\\hdashline[1pt/1pt]
\ra{Logit.io} & https://logit.io/\\\hdashline[1pt/1pt]
\ra{Lumigo} & https://lumigo.io/\\\hdashline[1pt/1pt]
\ra{New relic} & https://newrelic.com/\\\hdashline[1pt/1pt]
Ocelot & https://www.inspectit.rocks/\\\hdashline[1pt/1pt]
\ra{Opencensus} & https://opencensus.io/\\\hdashline[1pt/1pt]
\ra{Opentelemetry}& https://opentelemetry.io/\\\hdashline[1pt/1pt]
\ra{Sentry} & https://sentry.io/welcome/\\\hdashline[1pt/1pt]
Skywalking& https://skywalking.apache.org/\\\hdashline[1pt/1pt]
\ra{Site24x7}& https://www.site24x7.com/\\\hdashline[1pt/1pt]
\ra{Signoz}& https://signoz.io/\\\hdashline[1pt/1pt]
\ra{Splunk}& https://www.splunk.com/\\\hdashline[1pt/1pt]
Stagemonitor& https://www.stagemonitor.org/\\\hdashline[1pt/1pt]
Tanzu& https://tanzu.vmware.com/tanzu\\\hdashline[1pt/1pt]
\ra{Uptrace}& https://uptrace.dev/\\\hdashline[1pt/1pt]
\ra{Victoriametrics}& https://victoriametrics.com/\\\hdashline[1pt/1pt]
Zipkin& https://zipkin.io/\\\hdashline[1pt/1pt]

    \end{tabular}
\end{table}

\section{Tool analysis}
\label{sec:EmpiricalStudy}

\noindent The following steps---based on the research goal and derived questions described in Sect.~\ref{sec:intro}---describe the study context, the data extraction and analysis, and describe its verifiability and replicability following the approach suggested by \cite{WohlinExperimentation}.


\subsection{Study context}
We considered the eleven open tracing tools retrieved according the process described in Section~\ref{sec:ML}. Table~\ref{tab:Context} shows the selected tools with their \ra{respective} web site.

\subsection{Data extraction}
In this section, for each research question, we describe the collected data, i.e., the data we extracted from the retrieved search results.

\textbf{Distinctive features of each tool (RQ$_1$).} We extracted characteristics about the identified tools and grouped them into the following categories: 

\begin{enumerate}
    \item \textit{General information}: the links to the source code repository, the chosen licenses, the adopted programming languages, and the \ra{different price types};
    \item \textit{Deployment}: the components contained in each tool, also in comparison to the suggested architecture defined by the Open Application Performance Management (OpenAPM) initiative \cite{openapm} and their supposed deployment;
    \item \textit{Usage}: the suggested steps to use the tools, i.e., to setup data collection and to use the collected data;
    \item \textit{Data}: the actual data that can be collected with each tool, also compared to what the OpenTelemetry \cite{opentelemetry} standard suggests: traces, metrics, and logs;
    \item \textit{Interoperability}: aspects that are important to guarantee a high degree of operability: the availability of an API, support for OpenTelemetry \cite{opentelemetry}, and if self-hosting is possible.
\end{enumerate}




\textbf{Tool popularity (RQ$_2$).} We evaluated the popularity in terms of how much the tools are mentioned in public online sources. The following sources were investigated:

\begin{itemize}
    \item \textit{Peer-reviewed literature}: Using the same sources as in Section~\ref{sec:ML}, we investigated the popularity of each tool by applying the following search string on all fields including title, abstract, body, and references: \textbf{(``tool Name'' OR ``tool url'') AND (*opentracing* OR ``open tracing'')}. In the case of tools with different names, we considered all variants in the ``OR'' term. Two authors independently evaluated the relevance of each publication reported by Google Scholar and Scopus, so as to exclude papers not written in English, false positives, or from different domains. In case of disagreement, a third author provided his/her opinion
    \item \textit{Online media}: Using the same sources as in Section~\ref{sec:ML}, we collected eventual posts, tags, users, groups or websites pertaining to the tools. In particular, we searched the tools’ own communities, eventual groups present on LinkedIn and Google groups, as well as the number of appearances in commonly used communities and discussion forums such as: StackOverflow, Reddit, DZo\-ne, and Medium.
\end{itemize}











\vspace{2mm}
\textbf{Benefits and issues (RQ$_3$ and RQ$_4$)} To get the different opinions, especially on the advantages and problems of each tool, we extracted the corresponding content of the discussion threads from popular technology forums, including \textit{StackOverflow}, \textit{Medium} and \textit{DZone}. StackOverflow is the largest forum of technology-related questions and answers (Q\&As) for developers and tech-enthusiasts. Compared to StackOverflow, DZone is also one of the world’s largest online communities for developers, but focuses more on new tech-trends, e.g., DevOps, AI and big data, Microservices, etc. Furthermore, similar to DZone, Medium is also a well-known technology forum that provides tutorials and reviewing articles. Comparatively, articles on Medium are more common-reader-friendly and written in a non-technical style. These three platforms are the largest tech communities that can be considered covering a representative school of opinions described in different styles. 

Due to a different availability of APIs and different crawling policies of these three platforms, we applied different data crawling strategies for each platform accordingly:


\begin{itemize}
    \item \textit{StackOverflow}. We applied API-based content crawling using the StackExchange API to retrieve the questions and answers regarding each selected tracing tool. Specifically, we used the advanced search API\footnote{\url{https://api.stackexchange.com/docs/advanced-search}} to extract all the questions that contain the name of the tool in either the title or the body of the question, together with the according answers. Please note, due to the daily query limitation of the API, the \textit{pagesize} parameter was set to the maximum (i.e., 100 results shown per query) to minimize the crawling time. 
    \item \textit{Medium}: We adopted a manual crawling approach to respect Medium's policy of not allowing web crawling with tools like BeautifulSoup\footnote{\url{https://www.crummy.com/software/BeautifulSoup/}}. First, we searched the name of each tool and obtained all the articles about the tool. For each article, we manually copy/pasted the content into an individual text file named with the tool name and a sequence number.  
    \item \textit{DZone}: We adopted a hybrid crawling approach combing manual search and the use of BeautifulSoup. Different from Medium, Dzone allows crawling with BeautifulSoup within each individual article but not within the list of search results. Hence, we conducted a hybrid crawling strategy by manually collecting all article URLs for each tool and then automatically crawled the article content for each URL using BeautifulSoup.
\end{itemize}

\subsection{Data Analysis}
\label{sec:DataAnalysis}
\noindent In this Section, we report the data analysis protocol adopted to answer the research questions.

\textbf{Distinctive features (RQ$_1$).}  We inspected the documentation and the website of each tool and mapped the different features. Then, we created a Table that reports which features is available for each tool.

\textbf{Tool popularity (RQ$_2$).} We use the statistics from both scientific sources and media sources to interpret and compare the popularity of the selected tools.

\begin{figure}[ht]
\centering
serse  \includegraphics[width=.4\textwidth]{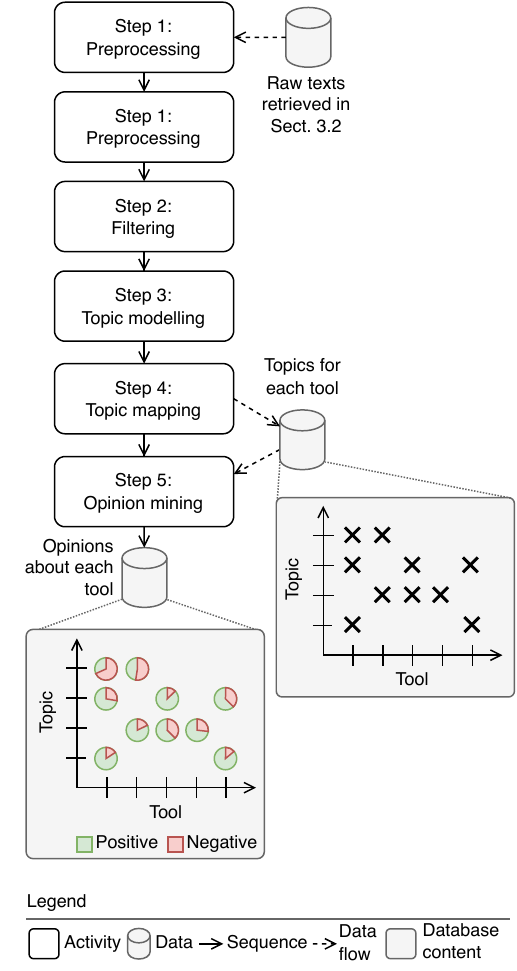}
  \caption{Approach to detect benefits and issues from analyzing social media texts (RQ$_3$ and RQ$_4$)}
  \label{fig:nlpframe}
\end{figure}

\textbf{Benefits and issues (\textbf{RQ$_3$} and \textbf{RQ$_4$}).} Figure \ref{fig:nlpframe} depicts the approach overview of how to elicit the benefits and the issues for the users regarding open tracing tools adoption based on the analysis of social media articles. 
\ra{For all the retrieved tools we crawled the gray literature data from the previously identified sources, and compare the total number of articles of all the tools. Due to the difference in their popularity (the answer to \textbf{RQ$_2$}), the retrieved textual data volume for each tool varies greatly. Sufficient data volume is highly necessary in order to obtain meaningful interpretation of practitioners' collective opinions. According to practitioners' experience, a potential minimum number of articles of 600 is required for Latent Dirichlet Allocation (LDA) modeling on news articles while 5k to 10k for tweets \cite{naushan2020topic}. In terms of the characteristics of the data for this study, we adopt the same threshold to verify the sufficiency and representativeness of the data. }



The rest of the process is composed by five main steps:

\begin{itemize}
    \item \textit{Step 1: Preprocessing}: This step pre-processes the raw text data and prepares them for further analysis. First, we divide texts from the dataset into sentence level instances since each text can contain multiple topics and various sentiments. Second, we build the bigram and trigram models, which means we identify the common phrases (e.g., \textit{New York} instead of \textit{new} and \textit{york}). Subsequently, for each sentence, a series of text processing activities are required, including transforming text into lower cases, removing non-alpha-numeric symbols, screening stop-words, and eliminating extra white spaces, and lemmatization.
    
    \item \textit{Step 2: Filtering}: This step is to filter out the non informative sentences with trained text classifier. By doing so, we shall identify the sentences that contain useful information and screen out those not relevant. Aiming to answer \textbf{RQ$_3$} and \textbf{RQ$_4$}, the informative sentences shall contain the explicit description on the benefits or issues regarding the tracing tools. For example, ``\textit{In fact, with automated instrumentation as part of \toolapp, metric data is produced consistently and comprehensively across all teams.}'' is informative by describing the benefit of using ; ``\textit{I checked the source code.}'' is non-informative and should be filtered out. 
    
    \item \textit{Step 3: Topic Modeling}: Herein, we detect the main topics of the informative sentences identified from the previous step. Using topic modeling techniques, we shall identify the aspects about which the articles are discussing. Especially, we use ChatGPT to support the effective summarization of each topic based on the according keywords.
    
    \item \textit{Step 4: Topic Mapping}: In this step, using the topic model built with the informative texts, we can map each piece of text, which is about one particular tool, to one or multiple topic. By doing so, we shall know regarding each tool which topics are discussed in the social media and how frequent for each topic. 
    
    \item \textit{Step 5: Opinion Mining}: Finally, using opinion mining techniques on the texts, we can also know the collective sentiment from the social media concerning each topic for each tool. Therefore, the outcome shall provide the detailed reflection on the percentage of positiveness and negativeness for each topic for each tool.
\end{itemize}


Based on the approach described above, the benefits of each tool can be obtained analyzing the detected topics in which the tool is discussed positively, based on the collected opinions, which shall answer \textbf{RQ$_3$}. Similarly, the issues of each tool can be summarized by the according topics in which the tool is mentioned negatively, which answers \textbf{RQ$_4$}.









\subsection{Verifiability and replicability}
\label{subsec:Replicability}
\noindent To allow our study to be replicated, we have published the complete raw data in the replication package.\footnote{\url{https://figshare.com/s/8aa40eea5d50ed27d347} \label{Package}}


\section{Results} 
\label{sec:Results}
In this Section, we report the obtained results by applying the steps described in the previous sections.

\subsection{Distinctive features (RQ\texorpdfstring{$_1$}{1})} 

To answer \textbf{RQ$_1$}, we studied each tool to extract its distinctive features. The results are as follows. 

\textbf{General information.} Table \ref{tab:ToolComparisonFeature}, for each tool summarizes the licenses we found in the repositories of the application, the reported programming languages, and the reference to the repository. We use the license identifiers defined by the SPDX workgroup \cite{spdx}. Please note that not all tools are entirely open: \toolapp, \tooldog, \toolint, \toollig, and \toolvmw~have proprietary parts (e.g., the backend) but release agents or libraries (see below) using an Open Source license. The reported programming languages for each tool are only taken from the repositories with an Open Source license. \ra{We also investigated the different pricing strategies of all the selected tools (shown in Table \ref{tab:ToolComparisonFeature}). Among the 30 retrieved tools, 12  are open-sourced fully and 6 of these 12 provide free version with limited features. Moreover, we reported the age of each tool by means of the year when the firs version was released on the market.}

\ra{To understand the level of support each tool can provide to a development team, Table \ref{tab:ToolComparisonFeature} lists the supported programming languages, but this is only one aspect. The results described below complete the picture: the identified architectural components listed in Table \ref{tab:architecture} help developers to understand how a tool needs to be integrated in their context; the type of collected metrics listed in Table \ref{tab:ToolComparisonMetrics} illustrate which type of metrics can be collected with the various tools, and Table \ref{tab:ToolComparisonInteroperability} compares the degree of interoperability each tool offers.}

\rb{Table \ref{tab:ToolComparisonFeature} lists those programming languages for which we found support in the relative repositories on GitHub and within the documentation. Those programming languages that are starred (*), support \textit{non-invasive} instrumentation, i.e., the automated modification of the code so that tracing information is sent to the Agent. Such non-invasive instrumentation is called in different ways by the producers, e.g., \tooldog\ calls it "Auto Instrumentation".}


\newcommand{\proela}{Go*, Python*, iOS*, Java*, NodeJS*, PHP*, Ruby*, Gherkin} 
\newcommand{\proinp}{Java*, JavaScript}
\newcommand{\projae}{Go*, Java*, NodeJS*, Python*, C++*, C\#*} 
\newcommand{\prosky}{Java*, Python*, NodeJS*, Lua*, JavaScript*, Rust*, PHP*}
\newcommand{\prosta}{Java*, HTML, JavaScript}
\newcommand{\prozip}{C\#*, Go*, Java*, JavaScript*, Ruby*, Scala*, PHP*}

\newcommand{\proapp}{Java*, Shell, .NET*, Python*, JavaScript, Go*, C/C++*, PHP*, NodeJS*}
\newcommand{\prodog}{Go*, Python*, Ruby*, JavaScript, Node.js*, Java*, .NET*}
\newcommand{\proint}{Shell, JavaScript, Go, Java*, Python*, .NET*, Clojure*, Kotlin*, Python*, PHP*, Scala*, NodeJS*, Ruby*}
\newcommand{\prolig}{Go*, JavaScript, Python*, Java*, HCL, .NET*, NodeJS*}
\newcommand{\provmw}{Java, Go, Python, JavaScript, Shell}

\newcommand{\repela}{https://github.com/elastic/apm-server}
\newcommand{\repoce}{https://github.com/inspectIT/inspectit-ocelot}
\newcommand{\repjae}{https://github.com/jaegertracing/jaeger}
\newcommand{\repsky}{https://github.com/apache/skywalking}
\newcommand{\repsta}{https://github.com/stagemonitor/stagemonitor}
\newcommand{\repzip}{https://github.com/openzipkin/zipkin}
\newcommand{\repapp}{https://github.com/Appdynamics} 
\newcommand{\repdog}{https://github.com/DataDog} 
\newcommand{\repint}{https://github.com/instana} 
\newcommand{\replig}{https://github.com/lightstep} 
\newcommand{\repvmw}{https://github.com/wavefrontHQ}

\begin{table*}
  \centering
  \footnotesize
  \setlength{\tabcolsep}{3pt}
  \caption{Features about each identified tool (RQ$_1$)}
  \label{tab:ToolComparisonFeature}
  \adjustbox{max width=\textwidth}{
  \begin{tabular}{l|L{3.5cm}L{4.3cm}L{0.7cm}L{4.2cm}L{1cm}}\hline
    \textbf{Tool} & \textbf{License} & \textbf{Programming language} & \textbf{Repo.} & \ra{\textbf{Pricing}} & \ra{\textbf{Created}}\\ \hline
    \ra{Appdash}& MIT \cite{repappdash} & Go*, Python*, Ruby* \cite{repappdash,langappdash} & \cite{repappdash} & Free & 2014\\\hdashline[1pt/1pt]   
    \toolapp & Proprietary, Apache-2.0, GPL-3.0, MIT~\cite{repapp} & \proapp~\cite{repapp,langapp} & \cite{repapp} & \$6/60/90/167 per month, per CPU Core; \$.06 per month, per 1000 tokens; Quote & 2008\\ \hdashline[1pt/1pt]
    \ra{Containiq}& Proprietary\cite{repcon} & C/C++*, Go*, Rust*, Python*, Ruby*, NodeJS* \cite{langcon} & \cite{repcon} &\$20 per node, per month OR \$.50 per GB of log data ingested; Quote & 2021\\\hdashline[1pt/1pt]
    \tooldog & Proprietary, Apache-2.0, BSD-3-Clause, GPL-2.0, MIT, MPL-2.0~\cite{repdog} & \prodog~\cite{repdog,autodog} & \cite{repdog} & Free; \$15/23 Per host, per month & 2010\\ \hdashline[1pt/1pt]
    \ra{Dynatrace} & Apache-2.0 \cite{repdyn}& C++*, .NET*, Erlang*, Go*, Java*, NodeJS*, Python*, Ruby*, Rust* \cite{repdyn,langdyn}& \cite{repdyn} & \$22/74/+15 per month for 8GB per Host; \$11 per month for 10K annual DEM Units; \$25 per month for 100k annual DDU; \$0.10 per CAU & 2005 \\\hdashline[1pt/1pt]
    \toolela & Apache-2.0, BSD-2-Clause, BSD-3-Clause, Elastic-2.0, MIT~\cite{repela} & \proela~\cite{repela,langela} & \cite{repela} & \$95/109/125/175 per month & 2012\\ \hdashline[1pt/1pt]
     \ra{Grafana tempo} & AGPL-3.0-only \cite{reptempo} & Java*, Go*, .NET*, Python*, NodeJS* \cite{reptempo,langtempo} & \cite{reptempo} & Free & 2020\\\hdashline[1pt/1pt]
    \ra{Haystack} & Apache-2.0 \cite{rephay} & Java*, NodeJS*, Python*, Go*, HCL, Shell, Smarty \cite{rephay,langhay} & \cite{rephay} & Free & 2017 \\\hdashline[1pt/1pt]
    \ra{Hypertrace}& Traceable Community License Agreement (1.0)\cite{rephyper}& Java*, Go*, Python*, NodeJs*, C++*, .NET* \cite{rephyper,langhyper} & \cite{rephyper} & Free & 2020 \\\hdashline[1pt/1pt]
    \ra{Honeycomb.io} & Apache-2.0, MIT\cite{rephoney}& Go*, Java*, .NET*, NodeJS*, Python*, Ruby*, JavaScript, Python \cite{rephoney,langhoney}& \cite{rephoney}& Free; Quote & 2016\\\hdashline[1pt/1pt]
    \toolint & Proprietary, Apache-2.0, GPL-2.0, MIT~\cite{repint} & \proint~\cite{repint,langint} & \cite{repint} & \$75/93.80 per host, per month & 2015 \\ \hdashline[1pt/1pt]
    \tooljae & Apache-2.0~\cite{repjae} & \projae~\cite{repjae,langjae} & \cite{repjae} & Free & 2016\\ \hdashline[1pt/1pt]
    \ra{Kamon}& Apache-2.0 \cite{repkamon} & Java*, Scala* \cite{repkamon,langkamon} & \cite{repkamon} & Free; \$89/299 per month; Quote & 2017\\\hdashline[1pt/1pt]
    \toollig & Proprietary, Apache-2.0, BSD-2-Clause, BSD-3-Clause, CC-BY-SA-4.0, MIT~\cite{replig} & \prolig~\cite{replig,langlig} & \cite{replig} & Free; \$100 per active service per month; Quote & 2015\\ \hdashline[1pt/1pt]
    \ra{Logit.io}& MIT \cite{replogit} & .NET*, Go*, NodeJS*, Python*, Ruby*, JavaScript, Shell \cite{replogit,langlogit}& \cite{replogit}& \$0.74 per GB; \$5 per million spans per month; \$2.80 per 1000 DPM & 2013\\\hdashline[1pt/1pt]
    \ra{Lumigo} & Apache-2.0\cite{replumigo}& Python*, NodeJS*, Java, Go \cite{replumigo,langlumigo}& \cite{replumigo}& \$99/299 per month; Quote & 2018 \\\hdashline[1pt/1pt]
    \ra{New Relic} & Apache-2.0 \cite{repnew} & C*, Go*, Java*, .NET*, NodeJS*, PHP*, Python*, Ruby*, JavaScript, Shell \cite{repnew,langnew} & \cite{repnew} & \$0.30 per GB Standard Ingest Cost Beyond Free Limits; + 0.50 per GB Data Plus Ingest Cost; +\$49 per month Core Users; +Quote & 2008\\\hdashline[1pt/1pt]
    \tooloce & Apache-2.0~\cite{repoce} & \proinp~\cite{repoce,langoce} & \cite{repoce} & Free & 2018\\ \hdashline[1pt/1pt]
    \ra{OpenCensus}& Apache-2.0 \cite{repopencensus} & Python*, NodeJS*, Go*, C\#*, C++*, Erlang*, Java* \cite{repopencensus,langopencensus} & \cite{repopencensus} & Free & 2017\\\hdashline[1pt/1pt]
    \ra{OpenTelemetry} & Apache-2.0 \cite{opentelemetry} & C++*, .NET*, Erlang*, Go*, Java*, JavaScript*, PHP*, Python*, Ruby*, Rust*, Swift* \cite{opentelemetry,langopentele} & \cite{opentelemetry} & Free  & 2019\\\hdashline[1pt/1pt]
    \ra{Sentry} & BSL-1.1 \cite{repsentry} & .NET*, JavaScript*, NodeJS*, Python*, PHP*, Rust*, Java*, Go* \cite{repsentry,langsentry} & \cite{repsentry} & \$0/26/80 per month; Quote & 2012\\\hdashline[1pt/1pt]
    \ra{Splunk} & Apache-2.0 \cite{repsplunk} & Python*, Java*, NodeJS*, .NET*, Go*, Ruby*, PHP* \cite{repsplunk,langsplunk}& \cite{repsplunk} & \$15 per host/month & 2003\\\hdashline[1pt/1pt]    
    \ra{Signoz}& MIT \cite{repsig}& Java*, Python*, JavaScript*, Go*, PHP*, .NET*, Ruby*, Elixir*, Rust* \cite{repsig,langsig}& \cite{repsig}& Free; \$200 per month; Quote & 2020 \\\hdashline[1pt/1pt]
    \ra{Site24x7}& BSD-2-Clause, MIT\cite{rep247}& Java*, .NET*, Ruby*, PHP*, NodeJS*, Python* \cite{rep247,auto247}& \cite{rep247}& €9/39/99/225 per month & 2006\\\hdashline[1pt/1pt]
    \toolsky & Apache-2.0~\cite{repsky} & \prosky~\cite{repsky,langsky} & \cite{repsky} & Free & 2015\\ \hdashline[1pt/1pt]
    \toolsta & Apache-2.0~\cite{repsta} & \prosta~\cite{repsta,langsta} & \cite{repsta} & Free & 2013\\ \hdashline[1pt/1pt]
    Tanzu& Apache-2.0 \cite{reptan}& Java*, C++*, Go*, .NET*, Python* Ruby \cite{reptan,autotan}& \cite{reptan}& Free & 2019\\\hdashline[1pt/1pt]
    \ra{Uptrace}& BSD-2-Clause, Apache-2.0 \cite{repupt} & Go*, NodeJS*, .NET*, Ruby*, Python* \cite{repupt,autoupt} & \cite{repupt} & \$0.10/0.09/0.08/0.07/0.06/0.05 per GB for 50/1000/2500/4250 /6000/8000GB with \$5+/100+ /200+/300+/400+/500+ budget & 2021\\\hdashline[1pt/1pt]
    \ra{Victoriametrics}& Apache-2.0 \cite{repvic}& Go*, JavaScript* \cite{repvic, autovic}& \cite{repvic} & Free; Quote & 2018\\\hdashline[1pt/1pt]
    \toolzip & Apache-2.0~\cite{repzip} & \prozip~\cite{repzip,langzip} & \cite{repzip} & Free & 2012\\ \hline
  \end{tabular}
}
\end{table*}

\begin{figure}[ht]
\centering
  \includegraphics[width=.32\textwidth]{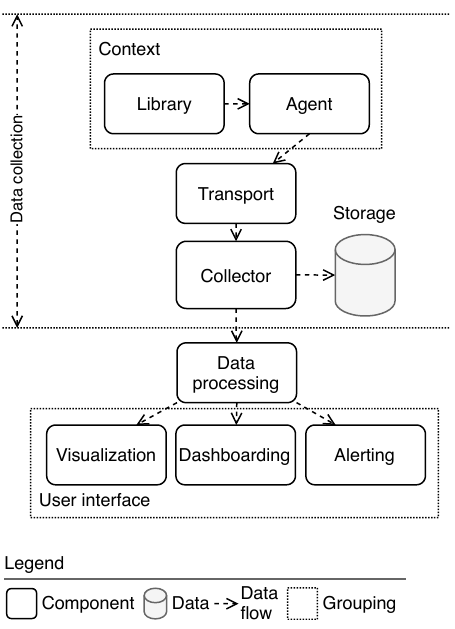}
  \caption{APM components according to the OpenAPM initiative \cite{openapm} and their typical communication data flow}
  \label{fig:architecture}
\end{figure}

\textbf{Deployment.} To better understand how each tool is supposed to be used in a tracing scenario, we studied the documentation of each tool to extract the suggested deployment configuration. 
However, before looking at how the various tools are deployed, it is useful to define the typical components of a tracing tool. We use the terminology defined by the Open Application Performance Management (OpenAPM) initiative \cite{openapm} (see Figure \ref{fig:architecture}): 

\begin{itemize}
    \item \textit{Libraries} are used in source code to send data to an agent or directly to the collection component. In some scenarios, agents are able to modify the application automatically so that it sends data to an agent without necessary source code changes (e.g., instrumenting Java byte code).
    
    \item \textit{Agents} are responsible for collecting data from a particular context, e.g., an application, the operating system, a mobile app, a database, or a web site. They usually run as part of applications or as an independent process and forward the data to collection components.

    \item \textit{Storage}: After the data is collected, it needs to be stored. To improve performance, this can occur through a \textit{transport} component that can fulfill routing or caching tasks. \textit{Collectors} receive data from agents or other data sources and persist it to \textit{Storage} components, e.g., a time-series database. 
    
    \item \textit{Data processing} components elaborate incoming data according to the analysis goals and prepare it for being used; the OpenAPM initiative distinguishes \textit{visualizations} (e.g., in form of charts), \textit{dashboarding}, and \textit{alerting}.
    
\end{itemize} 

Figure \ref{fig:architecture} depicts a generalized deployment scenario of the various components using the terminology of the OpenAPM initiative. The architecture depicted in Figure \ref{fig:architecture} also corresponds to the suggested architecture by the OpenTelemetry project~\cite{opentelemetry} (see below).

\ra{By studying the documentation of all the selected tools, we noted that only 7 out of the 30 tools (i.e., Dynatrace, Jaeger, New Relic, Sentry, Signoz, SkyWalking, and Tanzu) explicitly comply with the architecture described in Figure \ref{fig:architecture}. For majority of the tools, clear description of the ``Transport" features is missing. Meanwhile, a large majority of the tools contain their own UI while some tools, e.g., Ocelot, require visualization tools, e.g., Grafana to display the tracing outcomes. Furthermore, a majority of the tools also provide libraries, agents, collectors, storage and data processing capabilities. }


Table \ref{tab:architecture} lists each retrieved tool, its type, the identified components using the terminology suggested by the OpenAPM initiative, and the source (next to the name of the tool), where we obtained this information. When a tool uses a different term for a component, we mention this below the table. Please note that this table contains the components that are explicitly mentioned in the documentation. The absence of a component, e.g. an explicit transport component, does not mean that such a component does not exist in the platform but rather that this component might be contained in another component, e.g. the agent. Table~\ref{tab:architecture} also shows that the used terminology is not standardized and that different producers and teams call components in different ways.

\begin{table}
  \setlength{\tabcolsep}{3pt}
  \centering
  \footnotesize
  \caption{Identified architectural components}
  \label{tab:architecture}
  \begin{tabular}{l|lllllllll}\hline
     \textbf{Tool} & \rot{\textbf{Libraries}} & \rot{\textbf{Agent}} & \rot{\textbf{Transport}} & \rot{\textbf{Collector}} & \rot{\textbf{Storage}} & \rot{\textbf{Data processing~~}} & \rot{\textbf{UI}} \\ \hline
        \ra{Appdash} & & & \yes & \yes & \yes& & \yes \\\hdashline[1pt/1pt] 
    \toolapp~\cite{arcapp} & \yes[1]  & \yes     &          & \yes[2]  & \yes[2]  & \yes[2]  & \yes[2] \\\hdashline[1pt/1pt]
    \ra{Containiq}& & \yes & & & & & \yes \\\hdashline[1pt/1pt]
    \tooldog~\cite{arcdog} & \yes     & \yes     &          & \yes[3]  & \yes[3]  & \yes[3]  & \yes    \\\hdashline[1pt/1pt]
    \ra{Dynatrace} & \yes & \yes & \yes & \yes & \yes & \yes & \yes \\\hdashline[1pt/1pt]    
    \toolela~\cite{arcela} & \yes[4]  & \yes     &          & \yes[5]  & \yes     & \yes     & \yes    \\\hdashline[1pt/1pt]
    \ra{Grafana tempo} & \yes[\rsb{18}] & \yes[\rsb{18}] &  & \yes & \yes & & \yes\\\hdashline[1pt/1pt]
    \ra{Haystack} & \yes & \yes &  & \yes &  & & \yes \\\hdashline[1pt/1pt]
    \ra{Hypertrace} & \yes[7] & \yes & & \yes & & \yes & \yes \\\hdashline[1pt/1pt]
    \ra{Honeycomb.io} & \yes[7] & \yes &  & \yes & &  & \yes \\\hdashline[1pt/1pt]
    \toolint~\cite{arcint} & \yes[6]  & \yes     &          & \yes[3]  & \yes     & \yes     & \yes    \\\hdashline[1pt/1pt]
    \tooljae~\cite{arcjae} & \yes[7]  & \yes     & \yes     & \yes     & \yes     & \yes     & \yes    \\\hdashline[1pt/1pt]
    \ra{Kamon} & \yes & \yes & & \yes & &  & \yes\\\hdashline[1pt/1pt]
    \toollig~\cite{arclig} & \yes[7]  & \yes[8]  &          & \yes[9]  & \yes[9]  & \yes[9]  & \yes[9] \\\hdashline[1pt/1pt]
    \ra{Logit.io}& & & & & \yes & \yes & \yes \\\hdashline[1pt/1pt]
    \ra{Lumigo}& \yes & \yes &  & \yes &  & \yes & \yes \\\hdashline[1pt/1pt]
    \ra{New Relic} & \yes & \yes & \yes & \yes & \yes & \yes & \yes \\\hdashline[1pt/1pt]
    \tooloce~\cite{arcoce} &          & \yes                                                           \\\hdashline[1pt/1pt]
    \ra{OpenTelemetry} & \yes & \yes &  & \yes & & \yes &   \\\hdashline[1pt/1pt]
    \ra{Sentry} & \yes&  \yes & \yes & \yes & \yes  & \yes & \yes                \\\hdashline[1pt/1pt]
    \ra{Splunk} & \yes & \yes & & \yes & \yes & \yes & \yes   \\\hdashline[1pt/1pt] 
    \ra{Signoz}& \yes & \yes & \yes & \yes & \yes & \yes& \yes\\\hdashline[1pt/1pt]
    \ra{Site24x7}& \yes& \yes & \yes & \yes & & & \yes \\\hdashline[1pt/1pt]
    \toolsky~\cite{arcsky} & \yes[10] & \yes     & \yes     & \yes[11] & \yes     & \yes[12] & \yes    \\\hdashline[1pt/1pt]
    \toolsta~\cite{arcsta} &          & \yes     &          &          &          & \yes[13]           \\\hdashline[1pt/1pt]
    Tanzu~\cite{arcvmw} & \yes[1]  & \yes     & \yes[14] & \yes[15] & \yes     & \yes     & \yes     \\\hdashline[1pt/1pt]
    \ra{Uptrace}& \yes & & & \yes & \yes & & \yes \\\hdashline[1pt/1pt]
    \ra{Victoriametrics}& \yes& \yes& & & & & \\\hdashline[1pt/1pt]
    \toolzip~\cite{arczip} & \yes     & \yes[16] &          & \yes[17] & \yes[17] & \yes[17] \\\hline
    \multicolumn{10}{l}{\textsuperscript{1} called \textit{SDK}} \\
    \multicolumn{10}{l}{\textsuperscript{2} called \textit{Controller}} \\
    \multicolumn{10}{l}{\textsuperscript{3} called \textit{Backend}} \\
    \multicolumn{10}{l}{\textsuperscript{4} called \textit{Tracer API}} \\
    \multicolumn{10}{l}{\textsuperscript{5} called \textit{APM Integration}} \\
    \multicolumn{10}{L{8.3cm}}{\textsuperscript{6} depending on the technology to monitor, the documentation calls the\newline\hspace{5pt}data collection component \textit{library}, \textit{sensor}, \textit{tracing SDK}, or \textit{collector}} \\
    \multicolumn{10}{l}{\textsuperscript{7} relies on the APIs and SDKs provided by the OpenTelemetry project \cite{opentelemetry}}\\
    \multicolumn{10}{l}{\textsuperscript{8} called \textit{Microsatellites}} \\
    \multicolumn{10}{l}{\textsuperscript{9} called \textit{Engine}} \\
    \multicolumn{10}{l}{\textsuperscript{10} called \textit{Probes}} \\    
    \multicolumn{10}{l}{\textsuperscript{11} called \textit{Receiver cluster}} \\    
    \multicolumn{10}{l}{\textsuperscript{12} called \textit{Aggregator cluster}} \\    
    \multicolumn{10}{l}{\textsuperscript{13} called \textit{Widget}, only in web applications, for debugging purposes} \\    
    \multicolumn{10}{l}{\textsuperscript{14} called \textit{Proxy}} \\    
    \multicolumn{10}{l}{\textsuperscript{15} called \textit{Service}} \\ 
    \multicolumn{10}{l}{\textsuperscript{16} called \textit{Reporter}} \\ 
    \multicolumn{10}{l}{\textsuperscript{17} called \textit{Server}} \\ 
    \multicolumn{10}{l}{\textsuperscript{18} \rsb{relies on Grafana Agent}}
  \end{tabular}
\end{table}

\textbf{Usage.} Regarding the usage of the tracing tools, we studied  their installation and setup requirements, as described in the documentation.

All tools (except \tooloce~and \toolsta, which are Agents) are based on a similar setup, based on the suggested measurement architecture that tracing tools are based on (see Figure \ref{fig:architecture}). Therefore, using tracing tools always involves the following steps:

\begin{enumerate}
    \item \textit{Backend installation}: If the tool is installed on premise (as e.g., \toolela~\cite{staela}, \toolsky~\cite{stasky}, or \toolzip~\cite{stazip}): installing the tool following the documentation;
    \item \textit{Backend setup}: Preparing the backend to receive data: this step might involve creating an account for the organization (also called \textit{tenant} in \toolapp~\cite{staapp}), selecting a data collection site to respect privacy regulations (as for, e.g., \tooldog~\cite{stadog}) and setting up a project. In the case of \toolela, which uses a combination of tools within the backend, these tools have to be configured and connected with each other.
    \item \textit{Agent setup and application instrumentation}: All tools require the installation of agents and their configuration~\cite{staapp,stadog,staint,stalig,staela,stajae,stasky,stavmw,stazip2}. All tools offer a variety of agents that are able to either a) automatically instrument an application or b) allow developers to manually instrument it. \textit{Automatic instrumentation} means that the target application is modified in such a way that it logs and transmits the required data to the agent without manual work, \textit{manual instrumentation} means that the developer has to modify the code manually using the provided library to send what is needed to the agent. The agents have to be configured that they send the data to the backend, linking the data to a particular project. \tooloce~and \toolsta~are agents and require the configuration of a backend, e.g., InfluxDB\footnote{\url{https://www.influxdata.com}}.
    \item \textit{Data processing}: once the data collection is in place, the various tools (see Figure \ref{fig:architecture}) allow three different types of data processing: a) exploratory data analysis querying the collected data or visualizing it in charts b) pre-defining frequently needed queries and charts and storing and presenting them in form of dashboards c) pre-defining queries and defining thresholds to obtain alerts if certain conditions are met.
\end{enumerate}

\textbf{Data.} As mentioned in the introduction, distributed tracing aims to track requests as they flow through the services of a distributed system. Therefore, foremost, distributed tracing tools collect data about textit{traces}, i.e., how a request traverses different services. In addition, tracing tools often also collect \cite{opentelemetry} \textit{metrics} and \textit{logs}: metrics are measurements that describe the state of the observed system, e.g., the \texttt{memory} utilization at timestamp \texttt{2022-07-03T18:53:55Z} of micro\-ser\-vice \texttt{1}. Logs are messages that developers emit with their code to inform about important events, e.g., that the event \texttt{ItemDeleted} was initiated by user \texttt{7} and occurred with the time\-stamp \texttt{2022-07-03T18:53:55Z}. 

Table~\ref{tab:ToolComparisonMetrics} reports which of the three aspects---tracing, metrics, and logs---are collected by the analyzed tools. All tools collect traces, which is obvious as we are looking at tracing tools, and all tools except \ra{Appdash, Hypertrace, Signoz and } \toolzip~allow the additional collection of metrics. \ra{Regarding log data, only Hypertrace, Opencensus, Splunk, Victoriametrics and Zipkin do not provide log data}. These additional data is linked to the component in which the current trace was recorded and can be helpful when observing a trace. Next to each cross we provide the point in the documentation describing the presence of a particular data collection capability.


\begin{table}[ht]
  \setlength{\tabcolsep}{5pt}
  \centering
  \footnotesize
  \caption{Metrics collected by the identified tools (RQ$_1$)}
  \label{tab:ToolComparisonMetrics}
  \begin{tabular}{l|llll}\hline
     \textbf{Tool} & \rot{\textbf{Traces}} & \rot{\textbf{Metrics~~}} & \rot{\textbf{Logs}} \\ \hline
    \ra{Appdash}& \yes~\cite{repappdash} &  & \yes~\cite{repappdash}\\\hdashline[1pt/1pt]
    \toolapp & \yes~\cite{traapp} & \yes~\cite{metapp} & \yes~\cite{logapp}\\\hdashline[1pt/1pt]
   \ra{ Containiq}& \yes~\cite{tmlcon}& \yes~\cite{tmlcon} & \yes~\cite{tmlcon} \\\hdashline[1pt/1pt]
    \tooldog & \yes~\cite{tradog} & \yes~\cite{metdog} & \yes~\cite{logdog}\\\hdashline[1pt/1pt]
    \ra{Dynatrace} & \yes~\cite{tracedyn} & \yes~\cite{metricdyn} & \yes~\cite{logdyn}                    \\\hdashline[1pt/1pt]    
    \toolela & \yes~\cite{traela} & \yes~\cite{metela} & \yes~\cite{logela}\\\hdashline[1pt/1pt]
    \ra{Grafana tempo}& \yes~\cite{docgrafana}& \yes~\cite{docgrafana}&\yes~\cite{docgrafana} \\\hdashline[1pt/1pt]
   \ra{Honeycomb.io}& \yes~\cite{tracehoney}& \yes~\cite{metrichoney}&\yes~\cite{loghoney} \\\hdashline[1pt/1pt]
   \ra{ Hypertrace}& \yes~\cite{tracehyper} & &\\\hdashline[1pt/1pt]
    \ra{Haystack} & \yes~\cite{tracehay} & \yes~\cite{metrichay} & \yes~\cite{loghay}                   \\\hdashline[1pt/1pt]
    \toolint & \yes~\cite{traint} & \yes~\cite{metint} & \yes~\cite{logint}\\\hdashline[1pt/1pt]
    \tooljae & \yes~\cite{trajae} & \yes~\cite{metjae} & \yes~\cite{logjae}\\\hdashline[1pt/1pt]
    \ra{Kamon}& \yes~\cite{tracekamon}& \yes~\cite{metrickamon}& \yes~\cite{logkamon}\\\hdashline[1pt/1pt]
    \toollig & \yes~\cite{tralig} & \yes~\cite{metlig} & \yes~\cite{loglig}\\\hdashline[1pt/1pt]
    \ra{Logit.io}& \yes~\cite{tracelogit}&\yes~\cite{metriclogit} & \yes~\cite{loglogit}\\\hdashline[1pt/1pt]
    \ra{Lumigo}& \yes~\cite{tracelumigo}& \yes~\cite{metriclumigo}& \yes~\cite{loglumigo} \\\hdashline[1pt/1pt]
    \ra{New Relic} & \yes~\cite{tracenew} & \yes~\cite{metricnew} &  \yes~\cite{lognew}                  \\\hdashline[1pt/1pt]
    \tooloce & \yes~\cite{traoce} & \yes~\cite{metoce} & \yes~\cite{logoce}\\\hdashline[1pt/1pt]
    \ra{Opencensus} & \yes~\cite{traceopencensus}& \yes~\cite{metricopencensus} & \\\hdashline[1pt/1pt]
    \ra{OpenTelemetry} & \yes~\cite{traceopentele} & \yes~\cite{metricopentele} & \yes~\cite{logopentele}                   \\\hdashline[1pt/1pt]
    \ra{Sentry} & \yes~\cite{tracesen} & \yes~\cite{metricsen} & \yes~\cite{logsen}                   \\\hdashline[1pt/1pt]
    \ra{Splunk} & \yes~\cite{tracespl} & \yes~\cite{metricspl} &                    \\\hdashline[1pt/1pt]    
    \toolsky & \yes~\cite{trasky} & \yes~\cite{metsky} & \yes~\cite{logsky}\\\hdashline[1pt/1pt]
    \ra{Site24x7}& \yes~\cite{tracesite}& \yes~\cite{metricsite}&\yes~\cite{logsite}\\\hdashline[1pt/1pt]
    \ra{Signoz}& \yes~\cite{tracesig}& & \yes~\cite{logsig}\\\hdashline[1pt/1pt]
    \toolsta & \yes~\cite{repsta} & \yes~\cite{metsta} & \yes~\cite{logsta}\\\hdashline[1pt/1pt]
    Tanzu& \yes~\cite{travmw} & \yes~\cite{metvmw} & \yes~\cite{logvmw}\\\hdashline[1pt/1pt]
    \ra{Uptrace}& \yes \cite{traceupt} & \yes \cite{metricsupt}& \yes \cite{logupt}\\\hdashline[1pt/1pt]
    \ra{Victoriametrics}& \yes~\cite{tracevic}& \yes~\cite{metricvic}& \\\hdashline[1pt/1pt]
    \toolzip & \yes~\cite{repzip} \\\hline
  \end{tabular}
\end{table}

\textbf{Interoperability.} 
To evaluate interoperability, we looked at three aspects: the presence of a documented API, the support for OpenTelemetry \cite{opentelemetry}, and if it is possible to self-host the tool, i.e., to install everything locally. 

OpenTelemetry is a ``vendor-neutral open-source Observability framework for instrumenting, generating, collecting, and exporting telemetry data such as traces, metrics, logs \cite{opentelemetry}.'' We found that it is supported by all tools except \toolsta.

The results are reported in Table \ref{tab:ToolComparisonInteroperability}. The gray crosses indicate that self-hosting is implicitly possible because the entire tool is provided with an OpenSource license. Please note, that it might be complex to perform a local installation, but technically, it is possible.


\begin{table}[ht]
  \centering
  \footnotesize
  \caption{Features about interoperability (RQ$_1$)}
  \label{tab:ToolComparisonInteroperability}
  \begin{tabular}{l|lll}\hline
    \textbf{Tool} & \textbf{API} & \rot{\parbox{2cm}{\textbf{OpenTelemetry support}}} & \rot{\parbox{2cm}{\raggedright \textbf{Self-hosting}}} \\ \hline
    \ra{Appdash}& & \yes~\cite{repappdash} & \\\hdashline[1pt/1pt]
    \toolapp & \yes~\cite{apiapp} & \yes~\cite{oteapp} & \yes~\cite{preapp} \\\hdashline[1pt/1pt]
   \ra{Containiq}& & \yes~\cite{aoscon} & \yes~\cite{reqcon} \\\hdashline[1pt/1pt]
    \tooldog & \yes~\cite{apidog} & \yes~\cite{otedog} &                    \\\hdashline[1pt/1pt]
      \ra{Dynatrace} & \yes~\cite{apidyn} & \yes~\cite{opendyn} & \yes~\cite{hostdyn}                    \\\hdashline[1pt/1pt]    
    \toolela & \yes~\cite{apiela} & \yes~\cite{oteela} & \implied           \\\hdashline[1pt/1pt]
    \ra{Grafana tempo}& \yes~\cite{apigrafana}& \yes~\cite{opengrafana} &\\\hdashline[1pt/1pt]
   \ra{Honeycomb.io}& \yes~\cite{apihoney} & \yes~\cite{openhoney}&\\\hdashline[1pt/1pt]
   \ra{Hypertrace}& & \yes~\cite{openhyper} & \\\hdashline[1pt/1pt]
    \ra{Haystack} &  & \yes~\cite{openhay} &                    \\\hdashline[1pt/1pt]
    \toolint & \yes~\cite{apiint} & \yes~\cite{oteint} & \yes~\cite{preint} \\\hdashline[1pt/1pt]
    \tooljae & \yes~\cite{apijae} & \yes~\cite{otejae} & \implied           \\\hdashline[1pt/1pt]
    \ra{Kamon}& \yes~\cite{apikamon}& \yes~\cite{openkamon} & \\\hdashline[1pt/1pt]
    \toollig & \yes~\cite{tralig} & \yes~\cite{metlig} & \yes~\cite{loglig}\\\hdashline[1pt/1pt]
    \ra{Logit.io}& \yes~\cite{apilogit} & \yes~\cite{openlogit}& \\\hdashline[1pt/1pt]
    \ra{Lumigo}& & \yes~\cite{openlumigo}& \\\hdashline[1pt/1pt]
    \ra{New Relic} & \yes~\cite{apinew} & \yes~\cite{opennew} &                    \\\hdashline[1pt/1pt]
    \tooloce &                    & \yes~\cite{oteoce} & \implied           \\\hdashline[1pt/1pt]
    \ra{Opencensus} & \yes~\cite{apiopencensus}& \yes & \\\hdashline[1pt/1pt]
     \ra{OpenTelemetry} & \yes & \yes &                    \\\hdashline[1pt/1pt]
     \ra{Sentry} & \yes~\cite{apisen} & \yes~\cite{opensen} & \yes~\cite{hostsen}                   \\\hdashline[1pt/1pt]
    \ra{Splunk} & \yes~\cite{apispl} & \yes~\cite{tracesen} &                    \\\hdashline[1pt/1pt]    
    \toolsky & \yes~\cite{apisky} & \yes[2]~\cite{otesky} & \implied           \\\hdashline[1pt/1pt]
    \ra{Site24x7}& \yes~\cite{apisite} & \yes[3]~\cite{opensite} & \\\hdashline[1pt/1pt]
    \ra{Signoz}& & \yes~\cite{opensig} &\\\hdashline[1pt/1pt]
    \toolsta &                    &  \yes~\cite{repsta}& \implied           \\\hdashline[1pt/1pt]
    Tanzu& \yes~\cite{apivmw} & \yes~\cite{otevmw} &                    \\\hdashline[1pt/1pt]
    \ra{Uptrace}& & \yes~\cite{openupt} &\\\hdashline[1pt/1pt]
    \ra{Victoriametrics}& \yes~\cite{apivic} &\yes[4]~\cite{openvic}  & \\\hdashline[1pt/1pt]
    \toolzip & \yes~\cite{apizip} & \yes[1]            & \implied           \\\hline
    \multicolumn{4}{l}{\parbox{8cm}{\textsuperscript{1} OpenTelemetry data can be exported to Zipkin \cite{otezip}.}} \\ 
    \multicolumn{4}{l}{\parbox{8cm}{\textsuperscript{2} Metrics can be reported to the OpenTelemetry receiver or imported using OpenTelemetry exporter; Traces and logs are not supported.}} \\ 
    \multicolumn{4}{l}{\parbox{8cm}{\textsuperscript{3} Site24x7's support for OpenTelemetry is currently in development.}} \\ 
    \multicolumn{4}{l}{\parbox{8cm}{\textsuperscript{4} OpenTelemetry support is requested within an issue.}} \\ 
  \end{tabular}
\end{table}

From the point of view of interoperability, also the data provided in Table~\ref{tab:ToolComparisonFeature} can be of relevance: this table describes the used licenses of the tool and the used programming languages.

\subsection{Tool popularity (RQ\texorpdfstring{$_2$}{2})}
\textbf{Tool popularity based on peer-reviewed literature}. For each tool, we searched the Peer-reviewed publications that mentioned it. We found that only three Open Tracing Tools have been cited by more than 10 papers: \toolzip~29 times, \tooljae~18 times, and \toollig~10 times. The other considered tools have been cited less than 10 times. 

\begin{figure}[t]
\centering
  \includegraphics[width=.48\textwidth]{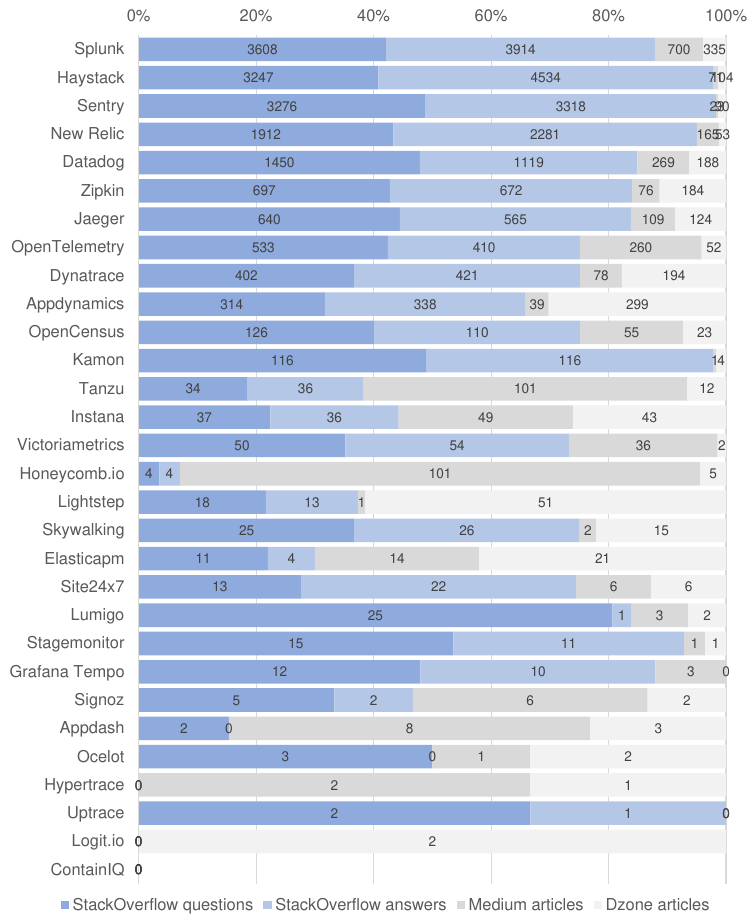}
  \caption{Social media content distribution (RQ$_2$)}
  \label{fig:smdist}
\end{figure}

\textbf{Tool popularity based on Online Media}. Shown in Figure \ref{fig:smdist}, the search results from three different technology-based social media platforms, i.e., StackOverflow, Medium, and DZone on each tool reflect their varied popularity. 

First of all, Splunk, among the \rb{30} tools, is the most popular considering the \rb{collective volume of the textual data from the three sources. The second and third most popular tools are Haystack and Sentry, which have similar levels of data volume compared to Splunk. New Relic and Datadog are also considerably popular and rank at 4th and 5th. However, the tools from 6th to 10th, i.e., Zipkin, Jaeger, OpenTelemetry, Dynatrace, and AppDynamics have only no more than 1/3 of data volume compared to the top ones. Furthermore, the bottom 20 tools have very limited social media coverage and have at best about 300 articles in all three sources. There are 14 tools having no more than 100 articles/posts in total. }

\begin{figure}[ht]
\centering
  \includegraphics[width=.48\textwidth]{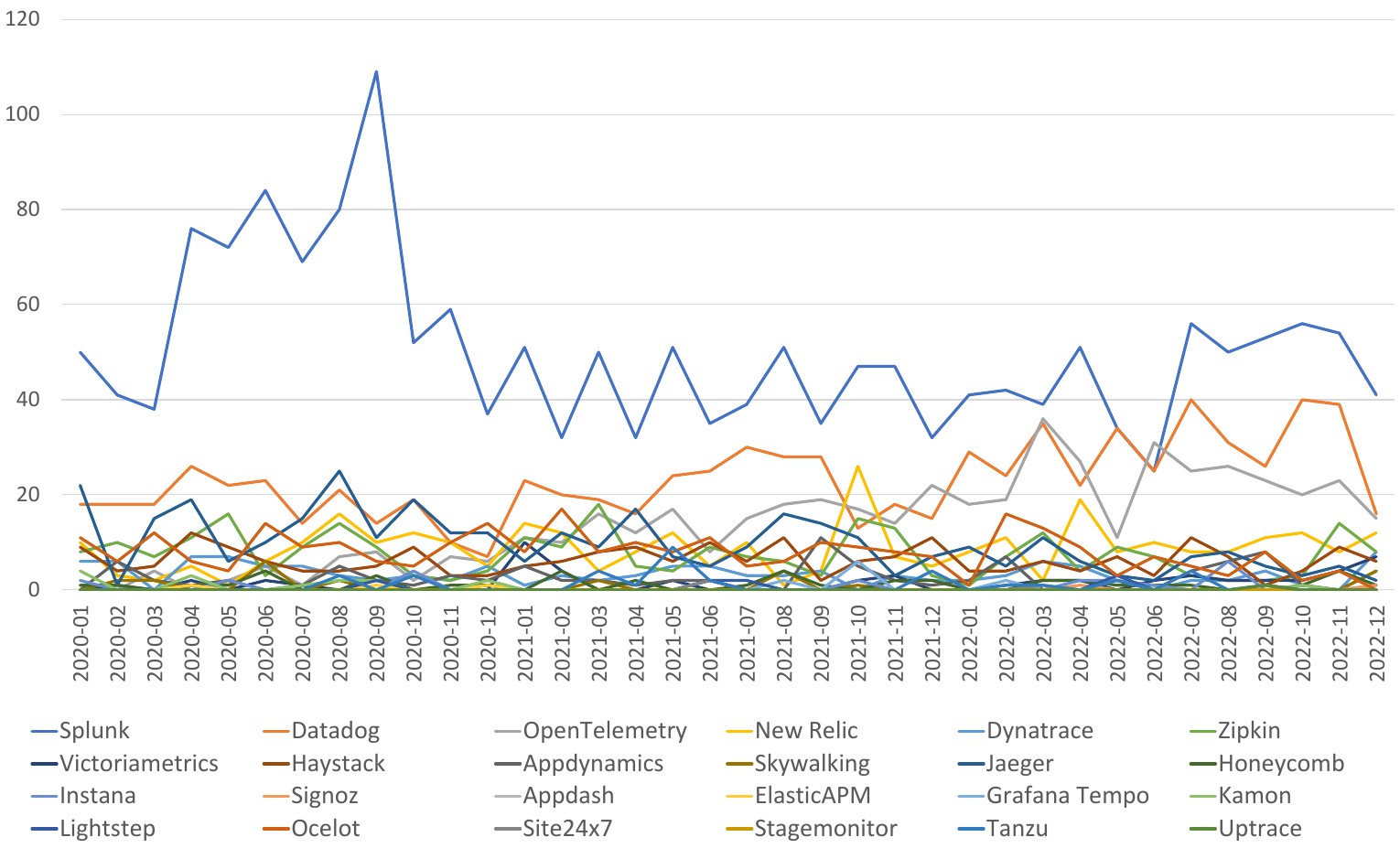}
  \caption{\ra{The Latest Discussion Frequency on Stack Overflow for each Tool}}
  \label{fig:soffreq}
\end{figure}

\rb{In Figure \ref{fig:smdist}, we reported the social media content distribution for the 30 retrieved tools. As we can see, 10 out of 30 take up 90.2\% of the articles of all the tools (Figure \ref{fig:smdist}). However, when checking the obtained data volume for each tool, we found only 10 tools have more than 600 data points \cite{naushan2020topic}. The discussion frequency (number of questions or answers each month) of each tool on Stack Overflow shows that only the selected tools have been drawing noticeable attention (shown in Figure \ref{fig:soffreq}). Unfortunately, 6 tools (i.e., ContainIQ, Hypertrace, Logit.io, Lumigo, OpenCensus and Sentry.io) did not provide available data points. The reason is likely due to the lack of discussion and their limited popularity.}

\subsection{Benefit and issues (RQ\texorpdfstring{$_3$}{3} and RQ\texorpdfstring{$_4$}{4})}

\ra{According to the social media content distribution described in Figure ~\ref{fig:smdist}, we proceeded to answer to  RQ$_3$ and RQ$_4$ only for the tools with sufficient data (as explained in Section~\ref{sec:DataAnalysis}). The final list is reported in Table \ref{tab:FinalList}. }  

\begin{table}[ht]
    \setlength{\tabcolsep}{4pt}
    \centering
    \footnotesize
    \caption{ \ra{The 10 tools considered for RQ$_3$ and RQ$_4$}}
    \label{tab:FinalList}
    \begin{tabular}{L{2.7cm}|l} \hline 
        \textbf{Tool name} &\textbf{Web site} \\ \hline 
        \ra{\toolapp} & https://www.AppDynamics.com \\\hdashline[1pt/1pt] 
        \ra{\tooldog} & https://www.Datadog hq.com/ \\\hdashline[1pt/1pt]
        \ra{\tooldyn} & https://www.Dynatrace.com/\\\hdashline[1pt/1pt]
        \ra{\toolhay} & https://www.Haystackteam.com/\\\hdashline[1pt/1pt]
        \ra{\tooljae} & https://www.Jaegertracing.io \\\hdashline[1pt/1pt]
        \ra{\toolnew} & https://opensource.newrelic.com/\\\hdashline[1pt/1pt]
        \ra{\toolope} & https://opentelemetry.io/\\\hdashline[1pt/1pt]
        \ra{\toolsen} & https://Sentry.io/\\\hdashline[1pt/1pt]
        \ra{\toolspl} & https://dev.Splunk.com/\\\hdashline[1pt/1pt]
        \ra{\toolzip} & https://zipkin.io \\  \hline 
    \end{tabular}
\end{table}

Following the approach described in Section~\ref{sec:EmpiricalStudy}, here we describe the obtained results for each step.

\textbf{Step 1: Preprocessing.} We pre-processed the crawled textual data by retaining only the natural language sentences. Herein we eliminated unnecessary content, such as the source code, URLs, publishing date and author info, etc. For Medium articles, we started eliminating the heading of the article that includes the publishing date and author info by splitting the string at the common last character of the part ``min read'' and selecting the later part. Subsequently, we used the sentence tokenizer from the Natural Language Toolkit (NLTK)\footnote{\url{http://www.nltk.org/}} to obtain the list of sentences from each article. As the tokenizer does not identify the source code or URLs, we eliminated them by selecting only the sentences ending with a period, an exclamation mark, or a question mark. 

First, we crawled data from social media, including, StackOverflow (\rb{16\,223} questions and \rb{17\,811} answers), Medium (\rb{2\,028} articles), and Dzone (\rb{1\,623} posts). \rb{We used \textit{langdetect} Python package~\footnote{\url{https://pypi.org/project/langdetect/}} to filter the non-English texts and obtained 37\,685 text data points, including \rb{17\,572} StackOverflow questions and \rb{16\,079} answers, \rb{1\,790} Medium articles and \rb{1\,623} Dzone posts. Furthermore, we filter the source code and html markdowns from each text using adapted \textit{html2text} package~\footnote{\url{https://pypi.org/project/html2text/}}. We obtained 338\,070 sentences using the NLTK sentence tokenizer (StackOverflow questions: 110\,524, answers: 65\,038, Medium: 65\,632, Dzone: 106\,876).  }

\begin{sloppypar}
\textbf{Step 2: Filtering.} Herein, we identified the informative sentences using a Na{\"i}ve Bayes (NB) classifier and the Expectation Maximization for Na{\"i}ve Bayes (EMNB) classifier \cite{nigam2000text}. The selection shall be based on the accuracy comparison of these two classifiers with the obtained dataset. First, we manually labeled a  sufficient number of training data including 50\% informative sentences and 50\% half non-informative ones. The selection criteria of informative sentences are that the sentence must explicitly present: 1) the benefits/features of the tools 
or 2) the issues of the tools. With an increasing number of training and testing data, the two classifiers shall be respectively trained and compared with the F1-score using a 5-fold cross validation. \end{sloppypar}

\begin{figure}[ht]
\centering
  \includegraphics[width=0.48\textwidth]{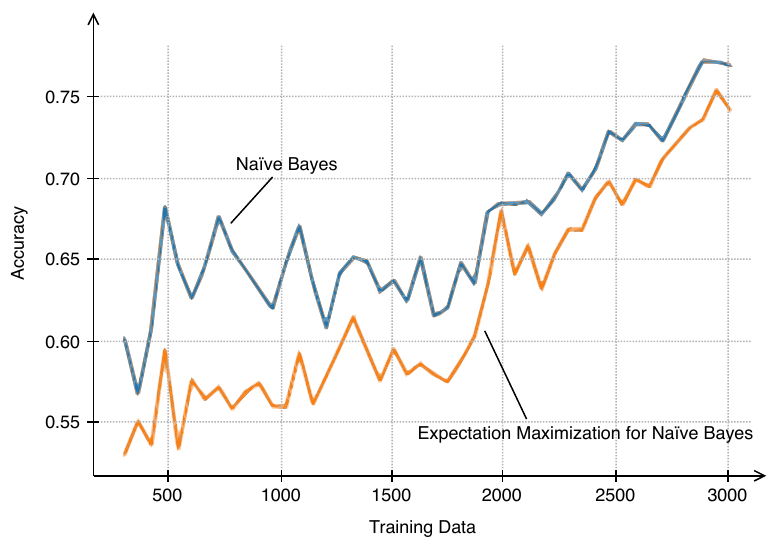}
  \caption{Testing informative text classifier accuracy (RQ$_3$ and RQ$_4$)}
  \label{fig:tclass_info}
\end{figure}

\begin{sloppypar}
In this study, from the \rb{338\,070} sentences obtained previously, we manually selected 3\,000 training data, including 1\,500 informative sentences and 1\,500 non-informative ones. 
To evaluate the performance of informativeness filtering, with a series of experiments, we compared the results of the NB algorithm and the EMNB algorithm with 3\,000 training data. We inspected the accuracy comparison of the two classifiers with different amounts of data starting from 200 to 3\,000 with an incremental step of 20. The test data ratio is set as default (0.25). Figure \ref{fig:tclass_info} shows that with the given training data, NB performs better than EMNB with the accuracy can reach as high as 0.76. Thus, we adopted the NB classifier for filtering the informative sentences. Using the classifier trained by the 3\,000 training data, we obtained \rb{158\,219} informative sentences. 
\end{sloppypar}

\textbf{Step 3: Topic Modeling.} To detect the topics of a set of text using an LDA topic modeling approach \cite{blei2003latent}, a number of preprocessing steps are required, which include: removing punctuation, removing extra space, restoring the word to its root form (lemmatization), remove stopwords, and build the bigram and trigram models. 

Furthermore, to find the best topic number for each review subset, we conducted a series of experiments for each set testing with the topic numbers ranging from 2 to 40. We used topic coherence to represent the quality of the topic models. Topic coherence measures the degree of semantic similarity between high scoring words in the topic. A high coherence score for a topic model indicates the detected topics are more interpretable. Thus, by finding the highest topic coherence score, we can decide the most fitting topic number. Herein, we use $c\_v$ coherence measure, which is based on a sliding window, one-set segmentation of the top words, and an indirect confirmation measure that uses normalized pointwise mutual information (NPMI) and the cosine similarity \cite{syed2017full}. Note that we pick the model that has the highest $c\_v$ value before flattening out or a major drop, in order to prevent the model from over-fitting. 

\begin{figure}[ht]
    \centering
    \includegraphics[width=0.48\textwidth]{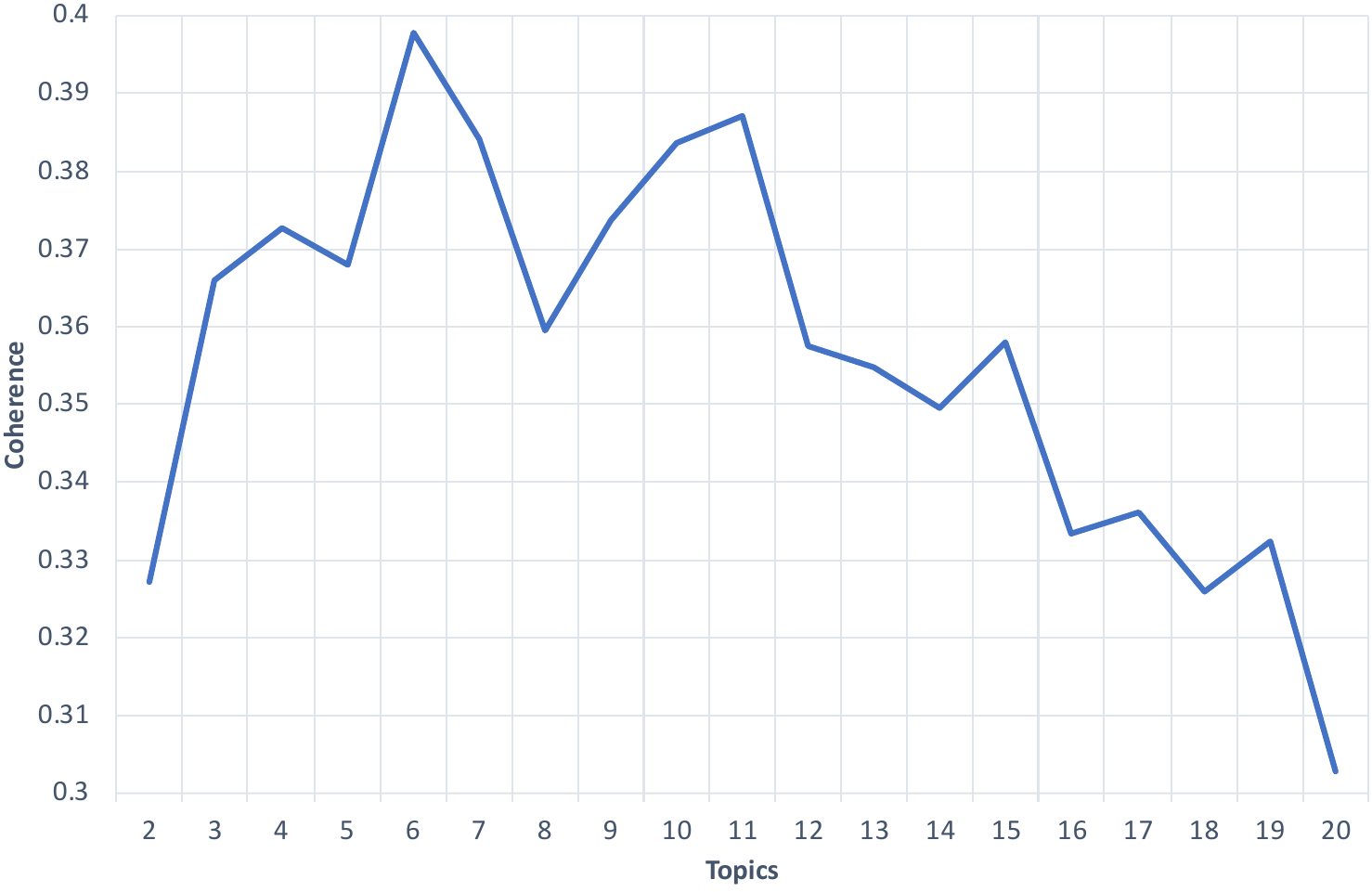}
    \caption{Topic number detection (RQ$_3$ and RQ$_4$)}
    \label{fig:topicnumber}
\end{figure}

Shown in Figure \ref{fig:topicnumber}, we built topic models using LDA with the number of topics from 2 to \rb{20} for text data. A clear turning point from the local highest value is at \rb{6. It is possible the coherence score reaches even higher when selecting topic numbers larger than 20. However, such a phenomenon is caused by the over-fitting models and shall be ignored. Thus, the topic number was determined as 6.}  

Therefore, with the LDA topic model, we detected the \rb{6} topics as follows: based on the allocated keywords in probability order, together with the overall term frequency as a reference, two domain experts synthesized their interpretation of the topics. The extracted topics are: \rb{ \textit{Usability}, \textit{Development}, \textit{Architecture}, \textit{Tracing}, \textit{Measurement}, \textit{Deployment \& Integration}. }

The list of topics and the according lists of indicator keywords are shown in Table \ref{tab:topickeywords}.

\begin{table}[ht]
    \setlength{\tabcolsep}{4pt}
    \centering
    \footnotesize   
    \caption{\ra{Topic interpretation with indicator keywords (RQ$_3$ and RQ$_4$)}}
    \label{tab:topickeywords}
    \begin{tabular}{L{2cm}|L{6.2cm}}
        \hline 
        \textbf{Topic} & \textbf{Indicator keywords} \\
        \hline
        \ra{Usability}&\ra{'user', 'support', 'developer', 'security', 'build', 'performance', 'production', 'design', 'use', 'need', etc.}\\\hdashline[1pt/1pt]
        \ra{Development}&\ra{'spring', 'team', 'software', 'development', 'implement', 'issue', 'develop', 'time', 'handle', 'process', etc.}\\\hdashline[1pt/1pt]
        \ra{Architecture}&\ra{'architecture', 'distribute', 'scale', 'observability', 'business', 'customer', 'solution', 'pattern', 'release', 'feature', etc.}\\\hdashline[1pt/1pt]
        \ra{Tracing}& \ra{'trace', 'source', 'code', 'framework', 'message', 'open', 'follow', 'alert', 'spring\_boot', 'transaction', etc.}\\\hdashline[1pt/1pt]
        \ra{Measurement}& \ra{'microservice', 'application', 'request', 'log', 'use', 'event', 'metric', 'server', 'kubernete', 'api', etc.}\\\hdashline[1pt/1pt]
        \ra{Deployment \& Integration}& \ra{'application', 'service', 'monitor', 'cloud', 'deploy', 'deployment', 'infrastructure', 'integration', 'manage', 'environment', etc.}\\
        \hline 
    \end{tabular}
\end{table}

\begin{figure}[]
\centering
  \includegraphics[width=.48\textwidth]{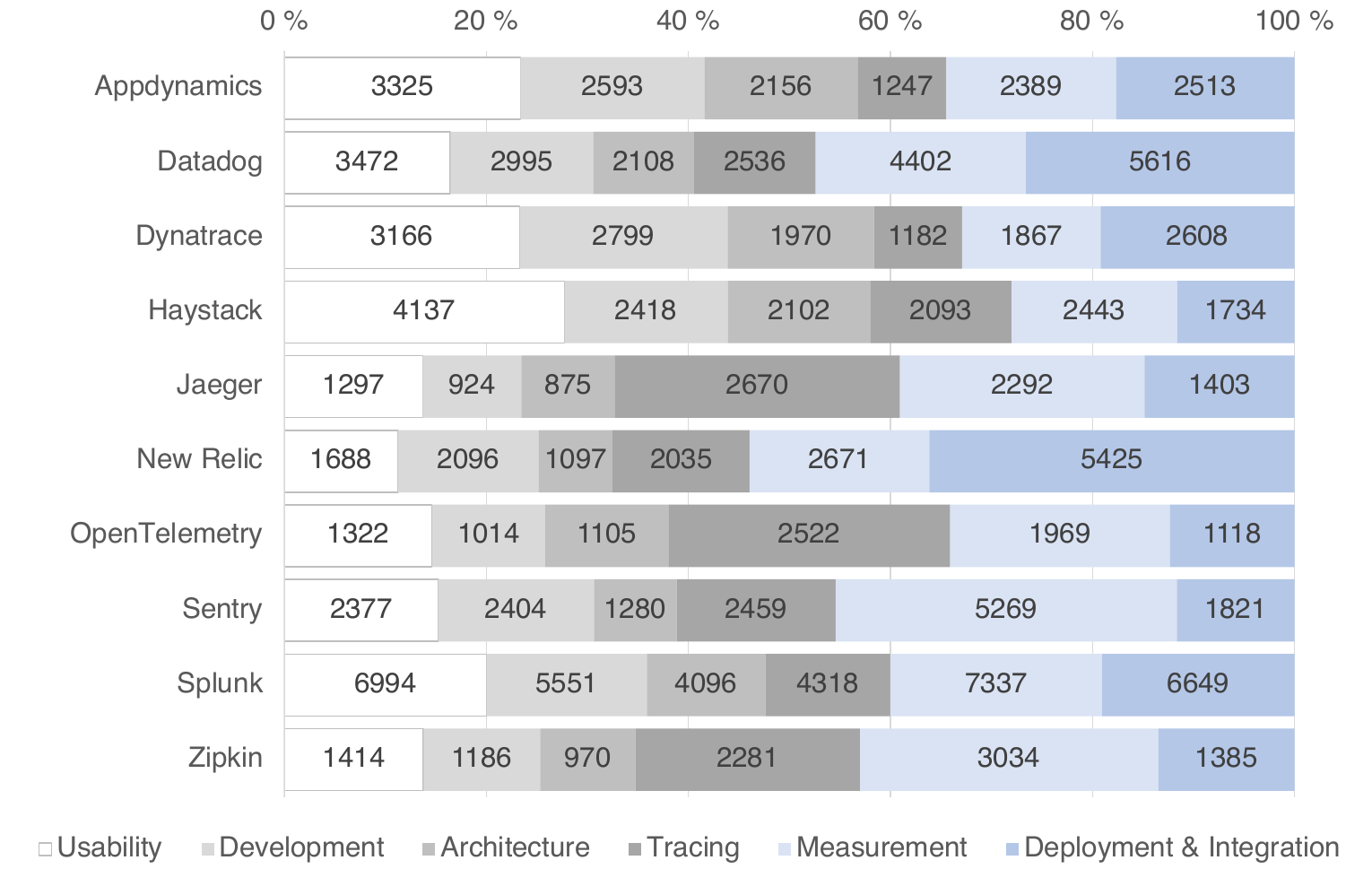}
  \caption{\ra{Topic frequency for each tool (RQ$_3$ and RQ$_4$) }}
  \label{fig:topicfreq}
\end{figure}

\textbf{Step 4: Topic Mapping.} With the obtained LDA topic mo\-del, we mapped each informative sentence to one of the topics to which it was most likely related. Shown in Figure \ref{fig:topicfreq}, the numbers of topic-related sentences from the articles on each tool are summarized. Compared with Figure \ref{fig:smdist}, the number of informative sentences for each tool correlated to that of the article crawled proportionally.  \rb{To be noted, we obtained much fewer informative sentences on Haystack than Splunk although they have a similar amount of article-level data points. The reason is that \textit{"needle in a Haystack"} is a classic algorithm problem metaphor for "checking a string contains another string", which has drawn heated discussion in StackOverflow. Meanwhile, "Haystack" can also be linked to the modular search for Django~\footnote{\url{http://Haystacksearch.org/}}. All such texts should have been classified as "non-informative". Displayed in Figure \ref{fig:topicfreq}, \textit{Deployment \& Integration} is the most dominant topic for Datadog  and New Relic. For some other tools, e.g., Haystack, Dynatrace and AppDynamics, \textit{Usability} topic is concerned the most.}

\begin{figure}[]
\centering
\begin{subfigure}[b]{102px}
    \includegraphics[trim=5 5 0 0,clip, height=190px, width=\textwidth]{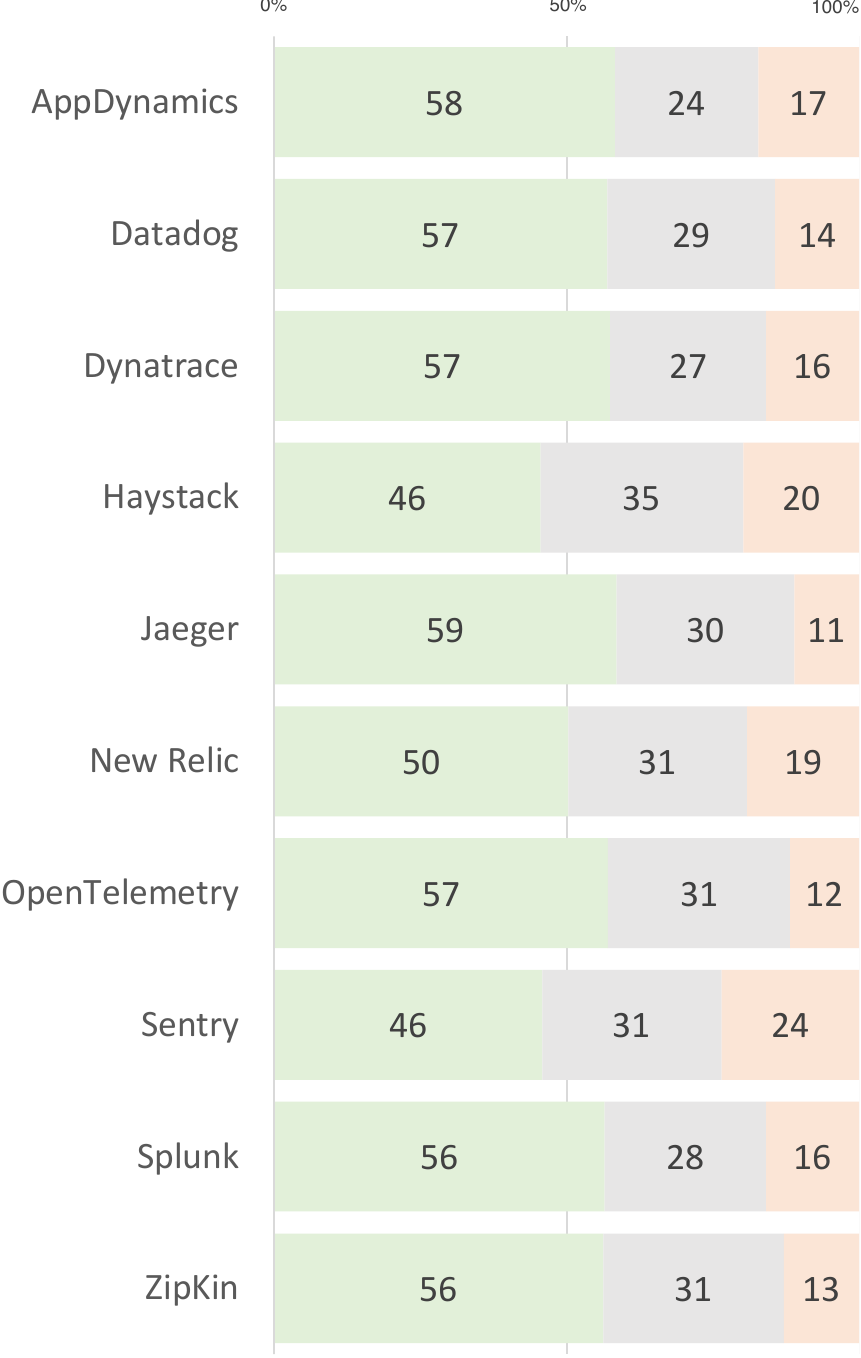}
    \caption{Usability}\hspace{0pt}
\end{subfigure}
\begin{subfigure}[b]{63px}
    \includegraphics[trim=118 5 0 0,clip, height=190px, width=\textwidth]{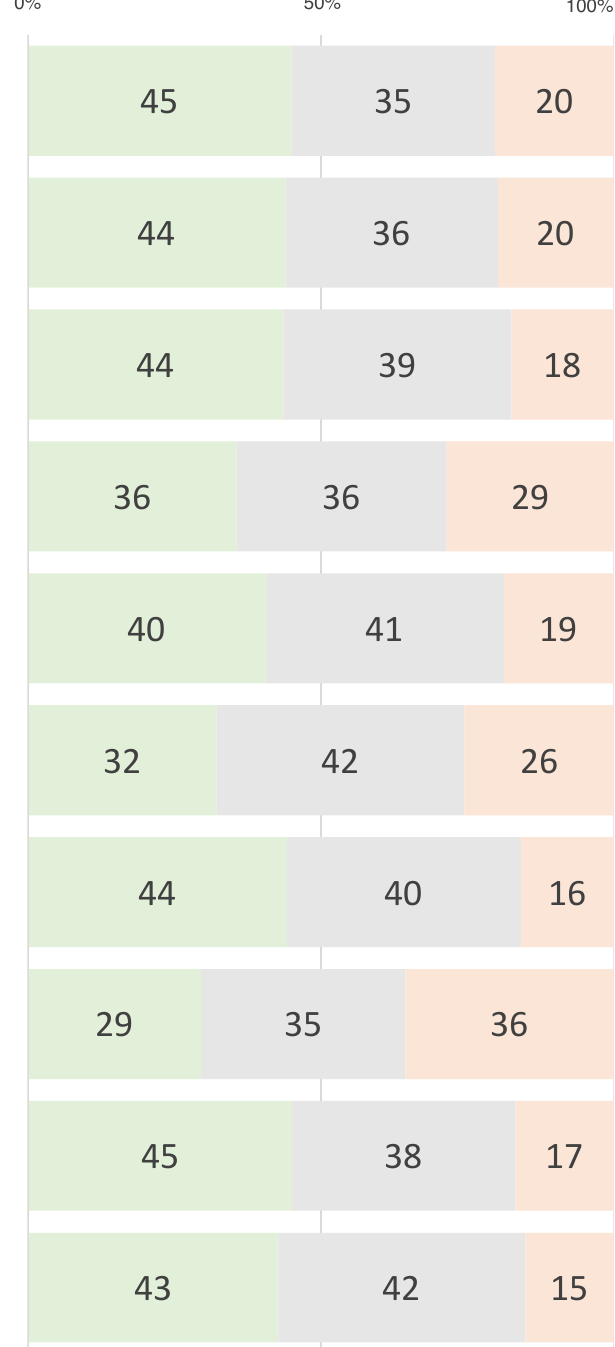}
    \caption{Development}\hspace{0pt}
\end{subfigure}
\begin{subfigure}[b]{63px}
    \includegraphics[trim=118 5 0 0,clip, height=190px, width=\textwidth]{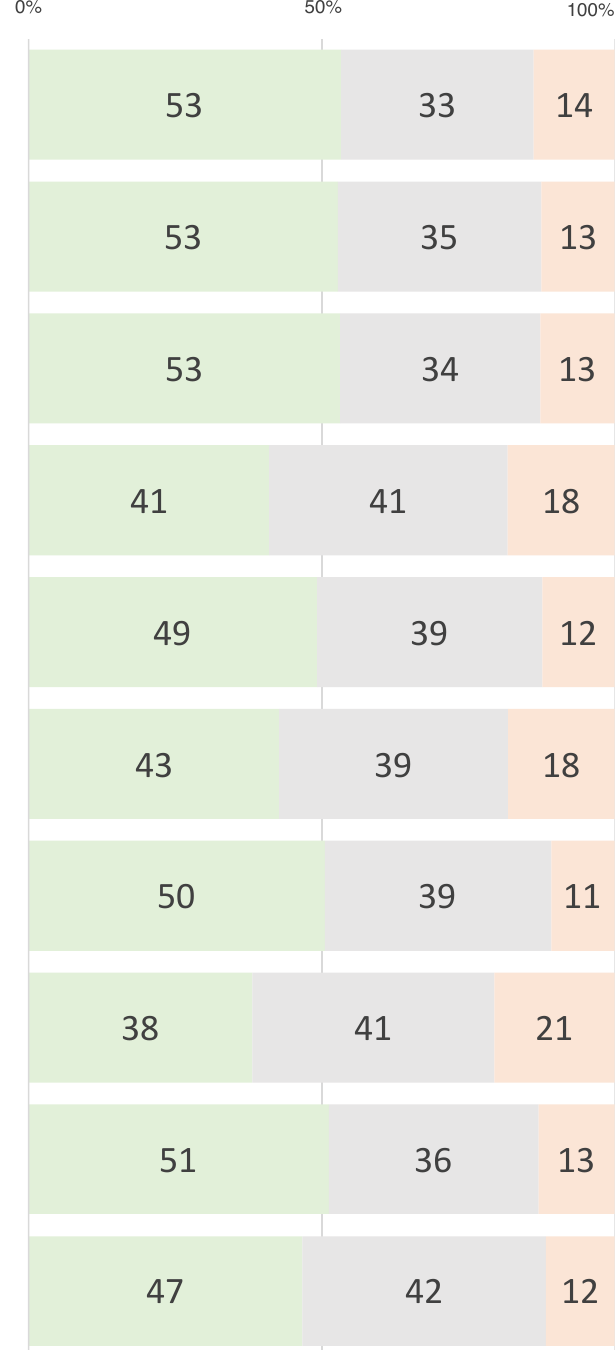}
    \caption{Architecture}\hspace{0pt}
\end{subfigure}

\begin{subfigure}[b]{102px}
    \includegraphics[trim=5 5 0 0,clip, height=190px, width=\textwidth]{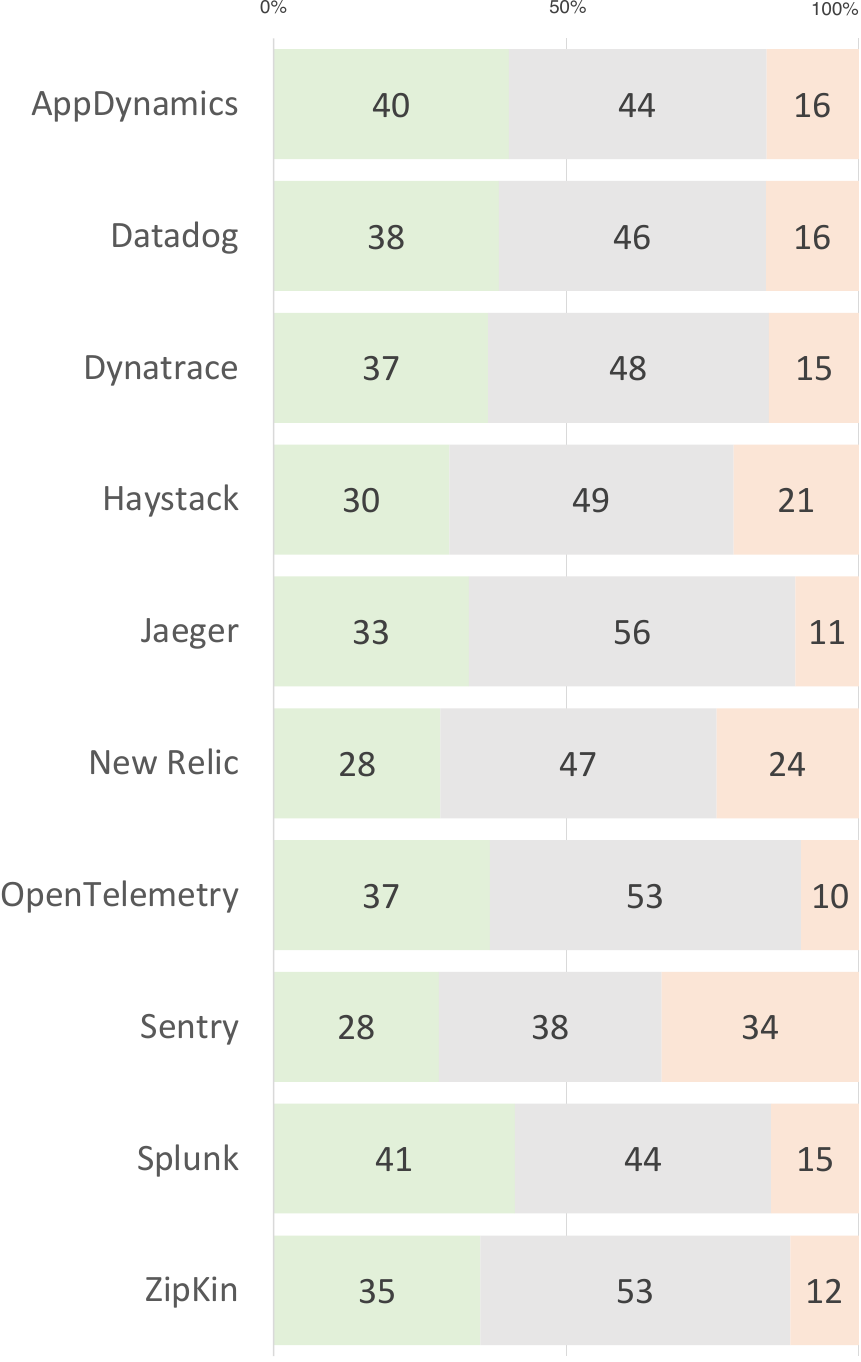}
    \caption{Tracing}\hspace{0pt}
\end{subfigure}
\begin{subfigure}[b]{63px}
    \includegraphics[trim=118 5 0 0,clip, height=190px, width=\textwidth]{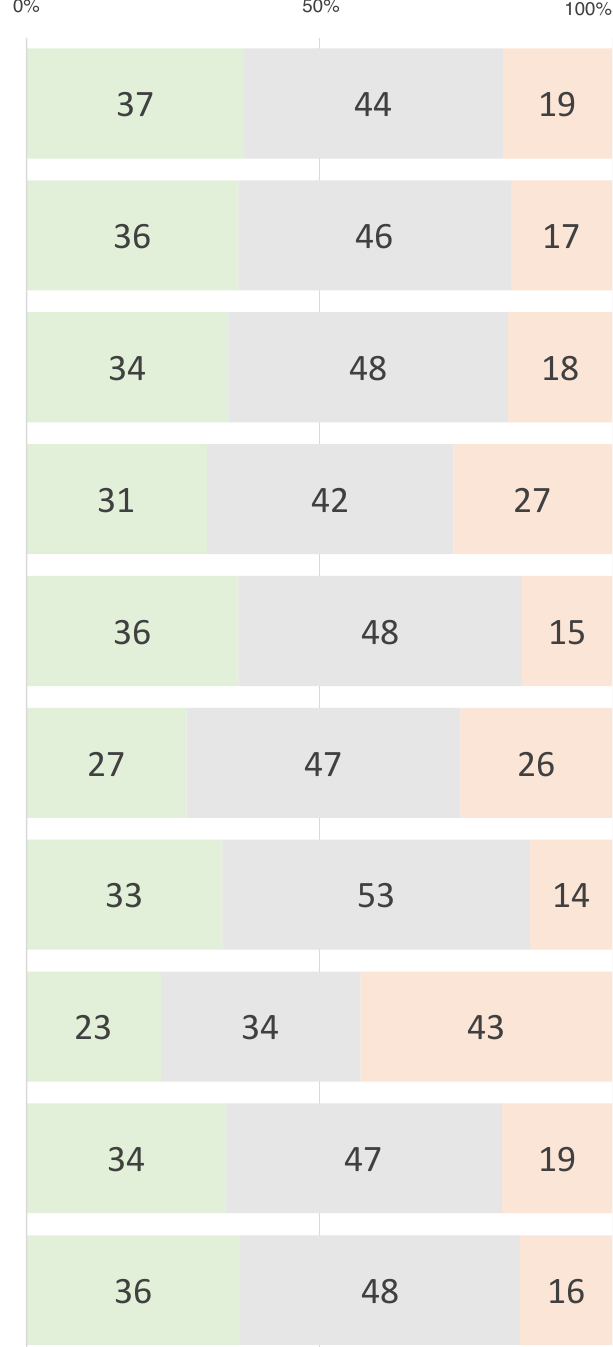}
    \caption{Measurement}\hspace{0pt}
\end{subfigure}
\begin{subfigure}[b]{63px}
    \includegraphics[trim=118 5 0 0,clip, height=190px, width=\textwidth]{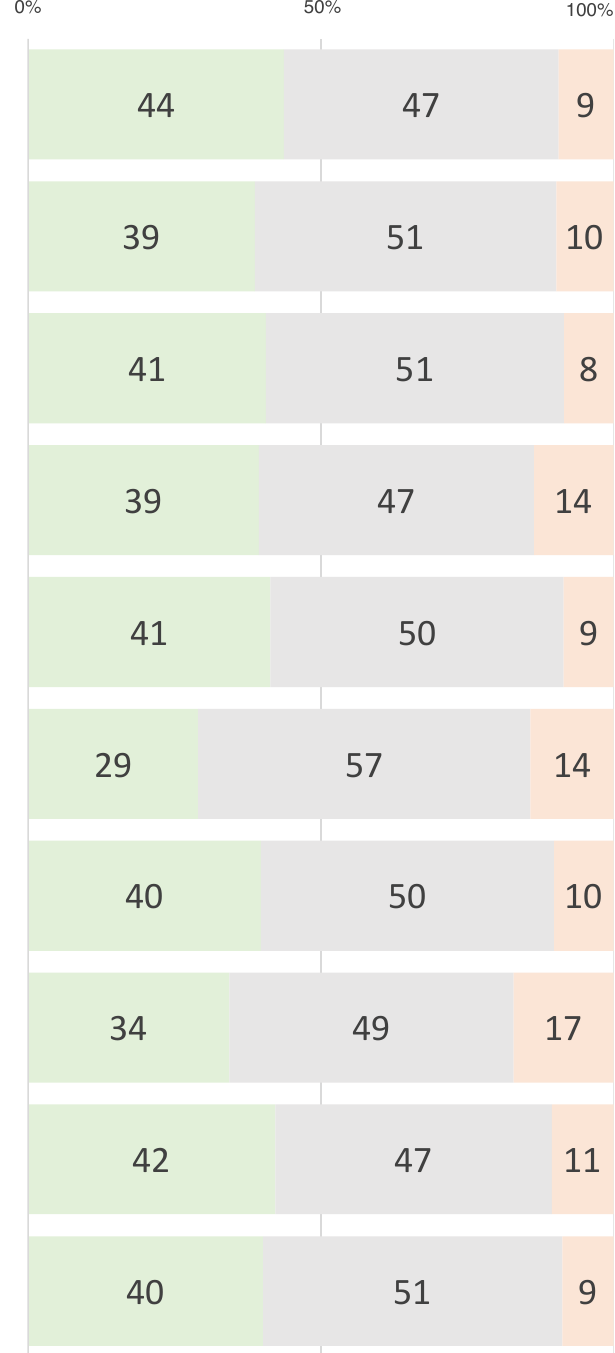}
    \caption{\centering Deployment \& Integration}
\end{subfigure}

\caption{\ra{Topic sentiment for each tool (RQ$_3$ and RQ$_4$)}}
\label{fig:topicsent}
\end{figure}

\textbf{Step 5: Opinion Mining.} Using the VADER method \cite{gilbert2014vader}, we can assess the sentiment of each informative sentence and furthermore the overall sentiment of each tool in terms of each topic. Herein, we take into account the percentage of positive, neutral, and negative sentences without considering the according sentiment strength. The percentage sentences in different sentiments for each tool on each topic is shown in Figure \ref{fig:topicsent}. To determine the benefits and issues in terms of each extracted aspect, i.e., topic, we compared each set of sentiment percentages to the average sentiment percentage of all sentences. 

Shown in Table \ref{tab:topicsenteachtool}, the percentage of the different sentiments for each tool was used as the reference. Therefore, we determined each topic for each tool being either a benefit, an issue, or a neutral opinion according to the following criteria.

\begin{table}[h]
    \setlength{\tabcolsep}{4pt}
    \centering
    \footnotesize
    \caption{\ra{Topic sentiment average percentage for each tool (RQ$_3$ and RQ$_4$)}}
    \label{tab:topicsenteachtool}
    \begin{tabular}{L{2.0cm}|rrr} \hline 
         \textbf{Tool}&\textbf{Positive}&\textbf{Neutral}&\textbf{Negative}\\ \hline 
        \ra{\toolapp} & \ra{47.4\%}& \ra{36.6\%}& \ra{16.0\%}\\\hdashline[1pt/1pt]
        \ra{\tooldog} & \ra{43.3\%}& \ra{42.2\%}& \ra{14.6\%}\\ \hdashline[1pt/1pt]
        \ra{\tooldyn} & \ra{45.7\%}& \ra{39.7\%}& \ra{14.6\%}\\ \hdashline[1pt/1pt]
        \ra{\toolhay} & \ra{38.0\%}& \ra{40.3\%}& \ra{21.7\%}\\ \hdashline[1pt/1pt]
        \ra{\tooljae} & \ra{40.8\%}& \ra{46.6\%}& \ra{12.6\%}\\ \hdashline[1pt/1pt]
        \ra{\toolnew} & \ra{32.5\%}& \ra{47.4\%}& \ra{20.1\%}\\ \hdashline[1pt/1pt]
        \ra{\toolope} & \ra{41.9\%}& \ra{46.2\%}& \ra{11.9\%}\\ \hdashline[1pt/1pt]
        \ra{\toolsen} & \ra{30.8\%}& \ra{36.6\%}& \ra{32.6\%}\\ \hdashline[1pt/1pt]
        \ra{\toolspl} & \ra{44.7\%}& \ra{40.1\%}& \ra{15.2\%}\\ \hdashline[1pt/1pt]
        \ra{\toolzip} & \ra{41.1\%}& \ra{45.8\%}& \ra{13.1\%}\\ \hline 
    \end{tabular}

\end{table}

\begin{itemize}
    \item If the percentage of positive sentences is higher than average and the percentage of negative ones lower than average, the topic is considered as a benefit for the tool.
    \item If the percentage of positive sentences is lower than average and the percentage of negative ones higher than average, the topic is considered as an issue for the tool.
    \item For any other circumstances, the topic is considered disputed.
\end{itemize}

Thus, according to the criteria, the according benefits and issues for each tool in terms of the topics can be summarized as Table \ref{tab:benefitissue}, which answers RQ$_3$ and RQ$_4$.

\begin{table}[]
    \setlength{\tabcolsep}{3pt}
    \centering
    \footnotesize
    \caption{\ra{Benefits and issues for each tool regarding topics (RQ$_3$ and RQ$_4$)}}
    \label{tab:benefitissue}
    \begin{tabular}{l|cccccccccc}
         \hline 
         \textbf{Criteria} &\rot{\textbf{\ra{\toolapp~~}}}&\rot{\textbf{\ra{\tooldog}}}&\rot{\textbf{\ra{\tooldyn}}}&\rot{\textbf{\ra{\toolhay}}}&\rot{\textbf{\ra{\tooljae}}}&\rot{\textbf{\ra{\toolnew}}}&\rot{\textbf{\ra{\toolope}}}&\rot{\textbf{\ra{\toolsen}}}&\rot{\textbf{\ra{\toolspl}}}&\rot{\textbf{\ra{\toolzip}}}  \\
         \hline 
 \ra{Architecture}&\benefit&\benefit&\benefit&\benefit&\benefit&\benefit&\benefit&\benefit&\benefit&\benefit\\\hdashline[1pt/1pt]

\ra{Deployment \& Integration}&&&&\benefit&\benefit&&&\benefit&&\\\hdashline[1pt/1pt]

 \ra{Development}&\issue&&\issue&\issue&\issue&\issue&&\issue&&\\\hdashline[1pt/1pt]

 \ra{Measurement}&\issue&\issue&\issue&\issue&\issue&\issue&\issue&\issue&\issue&\issue\\\hdashline[1pt/1pt]

 \ra{Tracing}&&\issue&\issue&&&\issue&&\issue&&\\\hdashline[1pt/1pt]

 \ra{Usability}&&\benefit&&\benefit&\benefit&\benefit&\benefit&\benefit&&\benefit\\\hdashline[1pt/1pt]
              

         
         
         
         
         
         
         \hline 
        \multicolumn{5}{l}{\benefit Benefit}\\
        \multicolumn{5}{l}{\issue Issue}
    \end{tabular}
\end{table}

\ra{We want to point out, that the "benefits" and "issues" obtained through the analysis of sentiments and discussion topics are complementary to the distinctive features of each tool (\textbf{RQ$_1$}): the \textit{actual} benefit or issue of using a tool in a given environment needs to be decided to know the context and the requirements of a given project. The benefits and issues identified in this section complement the picture and want to advise the practitioner that intends to use a given tool, to put particular care into the aspects we identified.}

\rsa{To further investigate the benefits and issues of each selected tool with details, we continued to use LDA topic modeling to extract the latent topics in each of the six criteria. For each subset of texts for each criterion of each tool (i.e., 6 topics $\times$ 10 tools = 60 subsets), We adopted the same ``topic-modeling-and-mapping" procedure as previously described. By conducting the same experiments to find the best topic number for each subset, we found the topic numbers (shown in Table \ref{tab:subtopicnumber} and trained the topic model for each subset. }

\begin{table}[ht]
    \setlength{\tabcolsep}{3pt}
    \centering
    \footnotesize
    \caption{\rsa{Sub-Topic Numbers for Each Topic of Each Tool}}
    \label{tab:subtopicnumber}
    \begin{tabular}{l|cccccccccc}
         \hline 
         \textbf{Criteria} &\rot{\textbf{\ra{\toolapp~~}}}&\rot{\textbf{\ra{\tooldog}}}&\rot{\textbf{\ra{\tooldyn}}}&\rot{\textbf{\ra{\toolhay}}}&\rot{\textbf{\ra{\tooljae}}}&\rot{\textbf{\ra{\toolnew}}}&\rot{\textbf{\ra{\toolope}}}&\rot{\textbf{\ra{\toolsen}}}&\rot{\textbf{\ra{\toolspl}}}&\rot{\textbf{\ra{\toolzip}}}  \\
         \hline 
 \ra{Architecture}&4&6&2&5&4&2&4&8&7&2\\\hdashline[1pt/1pt]

\ra{Deployment \& Integration}&3&3&6&6&2&4&5&2&3&6\\\hdashline[1pt/1pt]

\ra{Development}&7&5&2&2&4&4&4&8&2&7\\\hdashline[1pt/1pt]
 \ra{Measurement}&3&3&6&4&4&2&6&2&3&3\\\hdashline[1pt/1pt]
 \ra{Tracing}&3&9&5&4&4&3&3&3&4&2\\\hdashline[1pt/1pt]
 \ra{Usability}&9&4&3&4&3&4&4&8&3&6\\\hdashline[1pt/1pt] 

         \hline 
    \end{tabular}
\end{table}

\rsa{Same as the previous topic modeling step, we obtained a set of keywords sorted by the relevance for each of the 60 sub-topic. In addition, we used ChatGPT \cite{chatgpt} to support the effective interpretation and summarization of the topics based on these keywords. ChatGPT is an artificial intelligence chatbot developed by OpenAI, which uses foundational large language models (LLMs) and is fine-tuned via supervised and reinforcement learning techniques. Though a newly emerging technique, ChatGPT has quickly gained overwhelmingly world-wide attention from both industry and academia. Specifically, regarding the facilitation of text summarization, many early-stage studies have investigated the use of ChatGPT for such tasks \cite{yang2023exploring,luo2023chatgpt}. }

\rsa{Herein, for this topic extraction task, we adopted the newly released GPT-4 model\footnote{\url{https://openai.com/product/gpt-4}}. Compared to the legacy GPT-3.5 model, the new model has much higher reasoning capacity and conciseness. We initiate the topic extraction by entering a series of structured requests formatted as follows. }

\begin{center}
    \rsa{\emph{``Extract a short <MAIN TOPIC>-related topic for each of these lines of keywords.}}
    \\\emph{ }
    \\\rsa{\emph{<1st LINE OF KEYWORDS>}}
    \\\emph{...}
    \\\rsa{\emph{<$n$th LINE OF KEYWORDS>''}}
\end{center}

\rsa{The replies received were also structured, corresponding to the requests above as follows. }

\begin{center}
    \rsa{\emph{``<1st TOPIC>: <Explanation>}}
    \\\emph{...}
    \\\rsa{\emph{<$n$th TOPIC>: <Explanation>''}}
\end{center}

\rsa{For the purpose of validation, the first author and third author compare the AI-extracted topics and the original keyword lists. The sub-topics of each of the subsets are summarized in Table \ref{tab:subtopics} and Table \ref{tab:subtopics2}.}

\begin{table*}[]
  \centering
  \footnotesize
  \setlength{\tabcolsep}{3pt}
  \caption{\rsa{Sub-Topics for Each Topic of Each Tool (Part 1)}}
  \label{tab:subtopics}
  \adjustbox{max width=\textwidth}{
  \begin{tabular}{l|L{3.1cm}L{3.1cm}L{3.1cm}L{3.1cm}L{3.1cm}L{3.1cm}}
  \hline
  & \textbf{Usability} & \textbf{Development} & \textbf{Architecture} & \textbf{Tracing} & \textbf{Measurement} & \textbf{Deployment\&Integration} \\
  \hdashline[1pt/1pt] 
  \rotback{\textbf{AppDynamics}} & 1) \benefittopic{Tool production}; 2) User performance; 3) \issuetopic{App development pipeline}; 4) \benefittopic{Downtime reduction}; 5) \benefittopic{Security in application architecture}; 6) \issuetopic{Flexible data configuration}; 7) \issuetopic{Microservice implementation}; 8) Performance testing; 9) \benefittopic{Scalability and reliability}; & 1) \issuetopic{Microservice development}; 2) \benefittopic{Software testing processes}; 3) \issuetopic{Performance troubleshooting}; 4) \benefittopic{DevSecOps evolution}; 5) \benefittopic{Monitoring tools in DevOps}; 6) \issuetopic{Service integration and communication}; 7) \issuetopic{Reducing load times and costs}; & 1) \benefittopic{Microservice architecture patterns}; 2) \benefittopic{Mobile app customer experience}; 3) Cloud-based product architecture; 4) Observability in business applications; & 1) \benefittopic{Distributed tracing in microservices}; 2) Alerting and logging; 3) \issuetopic{Application performance and transaction tracing}; & 1) \issuetopic{API and microservice metrics}; 2) Measuring database and transaction performance; 3) \benefittopic{Application performance monitoring}; & 1) \benefittopic{Cloud and container deployment}; 2) \issuetopic{Application monitoring and integration}; 3) \benefittopic{Performance testing and continuous integration};\\
  \hdashline[1pt/1pt] 
  \rotback{\textbf{Datadog}} & 1) \benefittopic{Tool usability testing}; 2) \issuetopic{Performance monitoring and optimization}; 3) \issuetopic{Data pipeline management}; \benefittopic{Security and scalability in development environments}; & 1) \benefittopic{DevOps and Team Collaboration}; 2) Performance Issue Management; 3) \issuetopic{Optimizing Kubernetes Resource Usage}; 4) \issuetopic{Automation in Deployment Processes}; 5) Cloud-native Security Testing; & 1) \benefittopic{Microservices Architecture}; 2) \issuetopic{Latency Optimization in Distributed Tracing}; 3) API Reliability and Observability; 4) Scalability and Observability in Customer-focused Services; 5) Containerization and Enterprise Growth; 6) \issuetopic{Resource Management in High-Usage Environments} & 1) \benefittopic{Integrating Tracing Frameworks}; 2) Distributed Tracing 3) \issuetopic{Incident Notification and Response}; 4) Alert Configuration and Threshold Management; 5) Tracing in Containerized Environments; 6) \benefittopic{Open-source Event Tracing with Prometheus}; 7) Chaos Engineering and Observability; 8) \benefittopic{User Request Tracing and Performance Testing}; 9) \issuetopic{Error Tracing and Debugging in Kubernetes}; & 1) Kubernetes Metrics and Logging; 2) \benefittopic{Measuring Service Performance in Microservices}; 3) \issuetopic{Request-based Event Measurement}; & 1) \benefittopic{Cloud Deployment and Performance Monitoring}; 2) Chaos Engineering in Multi-layer Monitoring; 3) \issuetopic{Kubernetes Deployment and Datadog Integration}; \\
  \hdashline[1pt/1pt]
  \rotback{\textbf{Dynatrace}} & 1) \issuetopic{Tool use and user performance} 2) \benefittopic{Performance management in software development} 3) \benefittopic{Security and continuous development} &
1) \benefittopic{DevOps culture and collaboration} 2) \issuetopic{Proactive troubleshooting in software development} &
1) Cloud and open platform architectures 2) High-performance architecture for business applications &
1) Opentelemetry in observability 2) \issuetopic{Distributed tracing for optimizing response time} 3) Automated alerting and incident management 4) \benefittopic{Microservices and transaction tracing} 5) \benefittopic{Open-source tracing tools for performance analysis} &
1) \benefittopic{Performance measurement in microservices} 2) \issuetopic{Improving API performance with test metrics} 3) Monitoring and optimizing end-to-end service request time 4) Cross-platform performance measurement 5) \issuetopic{Log analysis and event rate limiting in distributed systems} 6) Enhancing app performance with load balancing and error management &
1) Deployment and monitoring of enterprise applications 2) Comparing application performance management (APM) tools 3) \benefittopic{Cloud-native application deployment and integration} 4) Load and performance testing in continuous integration 5) Dynatrace for telemetry and reporting 6) \issuetopic{Deployment and monitoring with Dynatrace} \\
  \hdashline[1pt/1pt]
  \rotback{\textbf{Haystack}} & 1) \benefittopic{Enhancing Customer Experience} 2) \issuetopic{Solving Usability Problems} 3) \benefittopic{Cybersecurity in Usability} 4) Implementing Search Indexes &
1) \benefittopic{Addressing Memory Leaks and Performance} 2) \issuetopic{Improving Error Handling and Search Functionality} &
1) Scalability in Cloud Architecture 2) Balancing Performance and Security in Enterprise Architecture 3) Enhancing Search Capabilities in Modern Applications 4) Customer-centric Architectures for AI-powered Applications 5) Embracing Microservices in Mobile Application Development &
1) \benefittopic{Visualizing User Interactions} 2) \issuetopic{Tracing Code Execution and Error Handling} 3) Integrating Elasticsearch and Haystack for Enhanced Search 4) Monitoring Email and Alert Response Times &
1) \benefittopic{Measuring API Performance} 2) \benefittopic{Evaluating Database Search Efficiency} 3) \issuetopic{Analyzing Server Performance and Error Impact} 4) \benefittopic{Monitoring Microservices and Application Metrics} &
1) Custom Service Deployment Challenges 2) \issuetopic{Integrating Elasticsearch with Django} 3) \benefittopic{Optimizing Query Performance} 4) \issuetopic{Validator Integration and Performance} 5) \benefittopic{Deploying and Integrating Search Functionality} 6) Monitoring and Scaling Deployment in Hadoop Environments \\
  \hdashline[1pt/1pt]
  \rotback{\textbf{Jaeger}} & 1) \issuetopic{Enhancing user and developer experience with microservices} 2) \benefittopic{Security and performance in application development} 3) Efficient software development & 1) \issuetopic{Memory management and CPU control in development environments} 2) \benefittopic{Streamlining the software development lifecycle} 3) \benefittopic{Building scalable, cloud-native applications} 4) Addressing challenges in software development & 1) \benefittopic{Balancing cost and performance in cloud-native architecture} 2) \issuetopic{Evaluating integration strategies and vendor solutions in architectural design} 3) \issuetopic{Achieving high availability and scalability in distributed data services} 4) \benefittopic{Enhancing observability in distributed microservice architectures} & 1) \issuetopic{Implementing Jaeger for performance monitoring and tracing} 2) Enhancing request tracking in applications 3) \benefittopic{Leveraging open-source tracing tools for distributed systems} 4) \issuetopic{Identifying and addressing errors in microservices} & 1) Analyzing end-to-end communication in service-based architectures 2) Evaluating application performance with log events and metrics 3) \benefittopic{Measuring Kubernetes-based microservices performance} 4) Assessing distributed service communication in microservice architectures & 1) \benefittopic{Streamlining cloud-native application deployment} 2) \issuetopic{Integrating and deploying Jaeger with Kubernetes for distributed tracing} \\
  \hline
  \end{tabular}
}
\end{table*}

\begin{table*}[]
  \centering
  \footnotesize
  \setlength{\tabcolsep}{3pt}
  \caption{\rsa{Sub-Topics for Each Topic of Each Tool (Part 2)}}
  \label{tab:subtopics2}
  \adjustbox{max width=\textwidth}{
  \begin{tabular}{l|L{3.1cm}L{3.1cm}L{3.1cm}L{3.1cm}L{3.1cm}L{3.1cm}}
  \hline
  & \textbf{Usability} & \textbf{Development} & \textbf{Architecture} & \textbf{Tracing} & \textbf{Measurement} & \textbf{Deployment\&Integration} \\
  \hdashline[1pt/1pt] 
  \rotback{\textbf{New Relic}} & 1) \benefittopic{Enhancing tool performance in applications} 2) \issuetopic{Simplifying error resolution} 3) \issuetopic{Streamlining problem-solving with New Relic} 4) Optimizing user experience in real-time production environments & 1) \issuetopic{Enhancing site performance and error handling} 2) Monitoring memory and CPU usage for server optimization 3) \benefittopic{Streamlining software development with New Relic} 4) \issuetopic{Identifying and resolving performance issues in applications} & 1) \issuetopic{Optimizing app architecture for high scalability} 2) \benefittopic{Implementing modern architecture with New Relic} & 1) \benefittopic{Enhancing function tracing with metrics} 2) \benefittopic{Leveraging transaction tracing for error detection} 3) \issuetopic{Analyzing page load times and tracing issues with New Relic} & 1) \benefittopic{Monitoring application performance with New Relic metrics} 2) \issuetopic{Identifying and resolving error causes through measurement} & 1) \benefittopic{Enhancing application performance with New Relic and Azure integration} 2) \benefittopic{Streamlining large-scale deployments with New Relic} 3) Ensuring smooth deployment through containerization and team collaboration 4) \issuetopic{Integrating New Relic agents for better monitoring and alerting} \\
  \hdashline[1pt/1pt]
  \rotback{\textbf{OpenTelemetry}} & 1) \issuetopic{Environment support and health} 2) Application performance and observability 3) End-user management in cloud architecture 4) \benefittopic{Open-source tools for resource optimization} & 1) \benefittopic{Cloud infrastructure development} 2) Adapting to changing needs in development 3) \benefittopic{Troubleshooting in microservice architecture} 4) \issuetopic{Optimizing time and cost in DevOps} & 1) \issuetopic{Navigating change in technology and deployment} 2) Scaling microservice architecture 3) Leveraging OpenTelemetry for enhanced observability 4) \benefittopic{Building distributed observability solutions} & 1) \issuetopic{Enhancing user experience with tracing} 2) \benefittopic{Adopting open-source tracing tools} 3) Integrating Jaeger and OpenTelemetry in data collection & 1) Monitoring distributed microservices 2) Utilizing OpenTelemetry exporters for data collection 3) Error management and Opentelemetry metrics 4) \benefittopic{Comprehensive metric collection in observability} 5) Enhancing database performance with tracing 6) \issuetopic{Optimizing database usage in microservices} & 1) \issuetopic{Zero-downtime deployment with Jaeger and OpenTelemetry} 2) \benefittopic{Open-source tools for application testing and deployment} 3) Streamlining cloud-native application management 4) Monitoring AWS services with custom metrics 5) \benefittopic{Adopting new monitoring platforms for enhanced observability} \\
  \hdashline[1pt/1pt]
  \rotback{\textbf{Sentry}} & 1) \benefittopic{Environment and Usability} 2) \benefittopic{Developer Tools} 3) \issuetopic{Error Resolution} 4) API Integration 5) \benefittopic{Data Management} 6) \benefittopic{Security Testing} 7) \benefittopic{Performance Optimization} 8) Scalable Security Solutions & 1) \benefittopic{Efficient App Development} 2) Scrum and Automation 3) \issuetopic{Error Handling in React Applications} 4) \issuetopic{Memory Management and Debugging} 5) Device-Optimized Development 6) \benefittopic{Scalable Infrastructure} 7) \benefittopic{Collaborative Development and Security} 8) \issuetopic{Continuous Integration and DevOps} & 1) \benefittopic{Adaptable Data Architecture} 2) Scalable and Resilient Table Architecture 3) \benefittopic{Community-driven Frameworks for Business Applications} 4) \issuetopic{Secure Data Management and Analytics} 5) Cloud-based App Development 6) Technology Selection and Tracking 7) Performance-driven Solution Development 8) \benefittopic{Automated Build and Release Processes} & 1) \benefittopic{Real-time Function Tracing} 2) \issuetopic{Error Monitoring and Communication} 3) \issuetopic{Open-source Tracing and Logging} & 1) \issuetopic{Error Measurement and Monitoring} 2) \benefittopic{Performance Metrics in API Services} & 1) \issuetopic{Seamless Application Deployment and Integration} 2) \benefittopic{Scalable Data Storage and Management} \\
  \hdashline[1pt/1pt]
  \rotback{\textbf{Splunk}} & 1) Streamlining Continuous Deployment 2) \issuetopic{Enhancing User Experience} 3) \benefittopic{Balancing Performance and Security} & 1) \issuetopic{Optimizing Application Performance} 2) \benefittopic{Advancing Agile DevOps} & 1) \issuetopic{Big Data and High-Performance Architecture} 2) Enhancing Observability in Software Architectures 3) Handling Massive Data Sets 4) Scalable Enterprise DevOps 5) \benefittopic{Agile Development for Complex Systems} 6) \issuetopic{Building Robust and Adaptable Applications} 7) \benefittopic{Optimizing Resource Estimation in Software Development} & 1) \issuetopic{Automated Alert and Notification Systems} 2) \issuetopic{Pinpointing Errors in Complex Systems} 3) Open Source Tracing in DevOps 4) \benefittopic{Real-time Data Visualization and Monitoring} & 1) Evaluating Test Efficiency 2) \benefittopic{Monitoring and Analysis of Microservices} 3) \issuetopic{Analyzing API and Service Performance} & 1) \issuetopic{Optimizing Test Performance in Deployment} 2) Adapting to Emerging Technologies and Custom Workloads 3) \benefittopic{Streamlining Cloud-Based Application Deployment and Integration} \\
  \hdashline[1pt/1pt]
  \rotback{\textbf{Zipkin}} & 1) \issuetopic{Ensuring a healthy build pipeline} 2) Organizational data ownership 3) \benefittopic{Scalability and reliability in service architecture} 4) Defining and measuring user metrics 5) \benefittopic{Microservice support tools} 6) \issuetopic{Managing production aspects in service environments} & 1) \benefittopic{Centralized monitoring tools} 2) \benefittopic{Exploring Spring framework and graph databases} 3) \benefittopic{Spring Boot for fast microservice development} 4) \issuetopic{Troubleshooting distributed microservices} 5) \issuetopic{Optimizing server and application performance} 6) Automating development and deployment in the cloud 7) \benefittopic{Cross-functional team collaboration in software development} & 1) \benefittopic{Observability in distributed microservice architecture} 2) \issuetopic{Scaling data-driven applications} & 1) \issuetopic{Error handling and tracing in service requests} 2) \benefittopic{Distributed tracing with Jaeger in Spring Boot applications} & 1) \issuetopic{Monitoring and measuring application performance} 2) Measuring microservice communication performance 3) \benefittopic{Assessing microservice scalability and resource usage} & 1) \benefittopic{Cloud-based microservice deployment and monitoring} 2) Integrating RESTful services in public and private cloud environments 3) Simplifying cloud application deployment with Spring Boot and JHipster 4) \issuetopic{Testing and monitoring microservices in production} 5) Enhancing service deployment with Istio and Envoy sidecars 6) \issuetopic{Kubernetes and Docker for streamlined microservice deployment} \\
  \hline
  \end{tabular}
}
\end{table*}

\subsubsection{RQ\texorpdfstring{$_3$}{3}. What benefits are achieved by adopting Open Tracing Tools?}


\rb{The results, summarized in Table \ref{tab:benefitissue}, show that all the selected tools provide benefits in terms of \textit{Architecture} based on the collective opinions of practitioners. On the other hand, 7 of the 10 tools, i.e., Datadog, Haystack, Jaeger, New Relic, OpenTelemetry, Sentry, and Zipkin, are positively received in terms of \textit{Usability}. Haystack, Jaeger, and Sentry also receive positive feedback regarding \textit{Deployment \& Integration}. However, in terms of \textit{Development}, \textit{Tracing}, and \textit{Measurement}, the opinions are more neutral or negative for all tools. To be emphasized, it does not mean none of these tools has any benefits for these aspects. It shows that the practitioners reflect more on their issues than the benefits regarding these aspects.}  

\rsa{Furthermore, by adopting another round of topic modeling, we further investigated each tool's collectively positive and negative opinions regarding each main topic. By doing so, we can intuitively compare each tool's benefits in more detail. The benefits of each tool summarized by the practitioners' collective opinions are shown in the green texts of Table \ref{tab:subtopics} and \ref{tab:subtopics2}}. 

\rsa{\textbf{Architecture}}. \rsa{Shown in the Table \ref{tab:benefitissue}, all tools are considered comparatively positively received, though some tools also received proportionally negative opinions on certain aspects. As shown in Table \ref{tab:subtopics} and Table \ref{tab:subtopics2}, AppDynamics is the only tool that has no significant issues. Especially, considering the keywords of the sub-topics, AppDynamics has the benefits regarding architecture patterns and addressing the challenges of scaling, security, and adoption in microservice-based business applications by leveraging design principles, cloud technologies, and containerization, as well as enhancing user experiences of mobile apps. Datadog is considered to perform well for microservice architecture in general and especially in tackling challenges and responsibilities in building distributed software systems with DevOps and monitoring tools. Comparatively, Jaeger is received more positively in terms of the balancing cost and performance in managing resources and complexity in large-scale enterprise deployments and also the enhanced observability in a distributed microservice architecture. Similarly, OpenTelemetry and Zipkin also have benefits in building distributed observability solutions. Sentry has benefits on adaptable data architecture in terms of developing flexible data models and storage solutions to accommodate changing customer needs and market demands while ensuring seamless access and integration for developers and users, community-driven frameworks for business applications towards developing scalable and adaptable business applications with a focus on observability, user experience, and seamless online integration, and automated build and release processes. Splunk is also received positively for the development process for complex systems, leveraging agile methodologies and cross-functional teams to effectively manage complex software projects and rapidly adapt to changing industry requirements, as well as the optimization of resource estimation in software development via employing data-driven models and analytics to improve project management, technology infrastructure, and resource allocation for on-time, cost-effective software delivery.}

\rsa{\textbf{Deployment \& Integration}}. \rsa{Regarding deployment and integration, many tools, e.g., AppDynamics, DataDog, Dynatrace, receive positive opinions on cloud and container deployment in general. Especially, according to the topic keywords, AppDynamics leans more towards the benefit of managing secure and efficient deployment of microservices in cloud-based and containerized infrastructure while Datadog focues on Streamlining application testing, deployment, and integration in scalable cloud environments to optimize the user experience. Dynatrace has the benefit of automating and managing containerized services with Kubernetes and continuous integration tools for secure, scalable, and efficient infrastructure management. Haystack has benefits in deploying and integrating search functionality. Jaeger, New Relic, and Splunk also have the benefits of streamlining deployment where Jaeger and New Relic are more praised for leveraging Kubernetes, containerization, and serverless architectures for efficient infrastructure management while New Relic for utilizing data from test reports, instance requests, and customer cases. On the other hand, OpenTelemetry is received more positively regarding leveraging Kubernetes, cloud vendors, and open-source resources for managing services and environments, while enhancing integration and traceability in development workflows, as well as  Leveraging telemetry data and diverse software tools for comprehensive infrastructure analysis while integrating with existing services and pipelines for seamless access and operation. }

\rsa{\textbf{Development}}. \rsa{Though shown in Table \ref{tab:benefitissue}, many tools are received negatively in terms of development, which does not mean there are no positive aspects at all. For example, several tools have benefits regarding DevOps and collaboration (Dynatrace and DataDog), DevSecOps evolution (AppDynamics), and Advancing Agile DevOps (Splunk). Specifically, based on the topic keywords, many specific aspects of DevOps are considered the benefits of these tools, including addressing security challenges, infrastructure management and cost optimization, automation tools, and processes etc. Sentry and Zipkin also have the benefit of collaborative development and security as well as scalability. Furthermore, for Zipkin, its benefits also include centralized monitoring tools, specifically concerning leveraging hardware and software solutions for monitoring traffic, alerting, and simplifying the management of logs and operational data in Azure-based applications, as well as the use of the Spring Boot platform to create efficient and scalable applications, with a focus on reactive data management and extending software design goals. }

\rsa{\textbf{Measurement}}. \rsa{Similarly, several benefits can also be found for each tool in the obtained sub-topics, though all tools are perceived relatively negatively regarding measurement. Surprisingly, for nearly all tools,  monitoring and measuring the performance of microservice architecture via application metrics, in general, is still positively perceived. Especially, OpenTelemetry also has benefits in the comprehensive metric collection in observability. Haystack and Sentry also have the benefits of measuring API performance. }

\rsa{\textbf{Tracing}}. \rsa{Regarding tracing, several tools have the benefits of distributed tracing (e.g., AppDynamics and Zipkin) and tracing for distributed systems (e.g., Jaeger). For these tools, using open-source tracing frameworks for monitoring and visualizing distributed microservices and enhancing observability and visualization are the benefits. On the other hand, real-time function tracing and real-time data visualization and monitoring are the benefits of Sentry and Splunk. Specifically, they enable leveraging free tools and alerts to trace function calls and state changes in real-time, optimizing query execution and exception handling, as well as interactive visualizations and tracking system performance in real-time enabling proactive incident response. }

\rsa{\textbf{Usability}}. \rsa{Regarding usability aspect, many tools are perceived more positively (shown in Table \ref{tab:benefitissue}. For AppDynamics, the benefits include tool production (enhancing developer productivity through feature-rich development platforms) downtime reduction (minimizing downtime, comparing and track system health), and scalability and reliability. Several tools have benefits regarding security, including AppDynamics, Datadog, Dynatrace, Haystack, Jaeger, Sentry and Splunk. The reason for security topics appearing under the usability aspect is likely due to the fact they are commonly mentioned simultaneously. On the other hand, DataDog has the benefit of tool usability testing (evaluating the effectiveness of feature design and user experience in application development) while Dynatrace also has the benefit of performance management (ensuring and improving user experience through end-to-end testing and performance monitoring in a DevOps business environment). Similarly, New Relic, Sentry, and Splunk are also perceived positively regarding tool performance optimization (implementing new usability features in monitoring tools to support end-user teams and improve customer experience through effective design and browser alerts) and management as well as the balancing between performance and security (leveraging tools and best practices in DevOps and risk management to maintain application security without sacrificing performance or customer satisfaction).}

\subsubsection{RQ\texorpdfstring{$_4$}{4}. Do Open Tracing Tools introduce any issues?}

Similarly, the issues introduced by each of the open tracing tools are also shown in Table \ref{tab:benefitissue}. \rb{We identify the issues from the tools that received more negative opinions than positive ones. Regarding \textit{Development}, AppDynamics, Dynatrace, Jaeger, Haystack, New Relic, and Sentry are perceived more negatively than the average. For \textit{Tracing} perspective, only Datadog, Dynatrace, New Relic, and Sentry are considered. Furthermore, all tools are perceived slightly negatively regarding \textit{Measurement}.} \rsa{Similarly, by adopting another round of topic modeling, we can further compare each tool's issues with more details. The issues of each tool summarized by the practitioners' collective opinions are shown in the red texts of Table \ref{tab:subtopics} and \ref{tab:subtopics2}. Apparently, as shown in the tables, it is likely there are still sub-topics that are perceived positively in those aspects. }

\rsa{\textbf{Architecture}}. \rsa{Datadog is poorly perceived regarding latency optimization in distributed tracing and resource management in high-usage environments. Jaeger has issues regarding evaluating integration strategies and vendor solutions and achieving high availability and scalability in distributed data services. New Relic and Zipkin also suffer from scalability-related issues regarding app architecture optimization and data-driven applications. For OpenTelemetry, the issue lies in navigating change in technology and deployment, that is, exploring the impact of release and deployment on observability, developer productivity, and key technology advancements while supporting user experience and tech leadership in API development. For Sentry, secure data management and analytics is the issue, that is, utilizing modern, open-source tools and APIs to develop secure, distributed data management systems with powerful analytics capabilities, ensuring efficient and safe handling of large-scale data across different platforms. Splunk's issues lie in big data and high-performance as well as robustness and adaptability.}

\rsa{\textbf{Deployment \& Integration }}. \rsa{Regarding deployment and integration, surprisingly, all tools have issues in deployment, integration and monitoring in general. Specifically, Jaeger has an issue regarding deployment with Kubernetes for distributed tracing. New Relic has issue in agents for monitoring and alerting. OpenTelemetry falls short regarding zero-downtime deployment when Sentry is perceived negatively regarding seamless application deployment. Zipkin has issues in streamlined microservice deployment and also in testing and monitoring microservices in production. }

\rsa{\textbf{Development}}. \rsa{Regarding the development aspect, AppDynamics is perceived negatively on microservice development (navigating the complexities of microservice applications to improve team productivity, implement AI-powered operations, and maintain a scalable and efficient development environment), performance troubleshooting (monitoring memory usage, CPU, and server metrics to identify, diagnose, and resolve performance issues), service integration and communication (efficient service integration, coordinated communication, and proactive issue resolution in complex engineering environments), and reducing load times and costs. Datadog has issues regarding Kubernetes resource optimization and deployment process automation. Dynatrace and Zipkin have issues in proactive troubleshooting when Haystack and New Relic are in error handling. Jaeger and Sentry are perceived negatively regarding memory management. OpenTelemetry, Splunk and Zipkin also suffer from performance optimization. }

\rsa{\textbf{Measurement}}. \rsa{In terms of measurement, AppDynamics suffers from negative opinions on API and microservice metrics, specifically about monitoring and aggregating API requests, user interactions, and error logs to optimize microservice performance, ensure reliability, and manage resource usage. Datadog has issues regarding request-based event measurement, that is, analyzing response times, errors, and resource usage in Azure and Prometheus to optimize endpoint performance and user experience. Dynatrace is perceived negatively mainly on API performance improvement with test metrics and log analysis and event rate limiting in distributed systems. Haystack, New Relic, and Sentry have issues in error management through measurement. Splunk also has issues in analyzing API and service performance, specifically, using real-time data, event logs, and query analysis to monitor and improve the response times and reliability of APIs and other service endpoints in complex systems. Zipkin has issues in monitoring and measuring application performance, analyzing logs and metrics at various levels to identify errors, and tracing events. }

\rsa{\textbf{Tracing}}. \rsa{Regarding the tracing aspect, error monitoring, and handling is a common issue for many tools, e.g., Datadog, Haystack, Jaeger, Sentry, and Zipkin. Meanwhile, both Datadog and Splunk have issues with incident notifications and automated alerts. AppDynamics has issues in performance and transaction tracing, i.e., capturing and analyzing transaction data to identify slow response times, execution bottlenecks, and other performance issues in business applications. Dynatrace and New Relic have issues in response time optimization and analysis. }

\rsa{\textbf{Usability}}.\rsa{Several tools have issues in app development and data pipeline management, including AppDynamics, Datadog, and Zipkin. The issues lie in the following aspects respectively: streamlining app development and release processes to create efficient, secure, and scalable applications using modern coding patterns and technologies; enhancing the usability of data pipelines with access controls, pattern compliance, and organization-wide sharing; complete audits and checks for crucial application deployment plans, with a focus on security and continuous production control. AppDynamics also has issues with flexible data configuration and microservice adoption toward system performance enhancement in general. Dynatrace, Haystack, Jaeger and Splunk are criticized for usability and user experience problems in general. New Relic has issues in simplifying error resolution and streamlining problem-solving. OpenTelemetry has issues in environment support and health, specifically about optimizing traceability in microservices for better team collaboration, ensuring security, and addressing problems through policy changes in the production pipeline, while managing costs and time. Sentry has issues with error resolution, specifically proactive error detection, and simplified solutions.}

\section{Discussion}
\label{sec:Discussion}
The analysis of the results revealed interesting insights that let us distill a number of lessons and/or implications both for researchers and practitioners. 

\textbf{How to select a tool?} Based on the comparative results, we cannot conclude that any of these tracing tool candidates is clearly better than the others. According to the results obtained in our study, and specifically for \textbf{RQ$_1$}, different Open tracing tools provide different features that suit users and organizations with different preferences. For example, users who value complete control over their data will more likely choose a self-hosted solution; users who prefer a commercial solution with commercial support, will choose a tool like \toolapp~or \tooldog. Most of the time, the availability of an agent or library for the programming language used within the team will be a important criteria to choose a specific tool.

\rb{Please note that questions about run-time behavior like throughput, resource usage, and performance impact are also of concern for engineers, but were not discussed in this paper. The actual run-time behaviour of a given tool depends on the adopted programming language, the context in which the observed software is deployed, and on the level of granularity with which data is collected. We therefore suggest to \textit{first} select tools based on the required functionality and \textit{then} to validate in the concrete context, which tool satisfies more requirements.}

We compared the \ra{30} selected open tracing tools in details regarding licensing, programming languages, deployment, usage, the collected data, and interoperability. We hope that these results will ease the effort of the teams in selecting a tool according to their needs. 

\begin{sloppypar}
\textbf{Which tools are best for what?} The outcome of opinion mining from the gray literature (\textbf{RQ$_3$} and \textbf{RQ$_4$}) shows the tools' quality reflected by the practitioners in the \ra{six} key aspects \ra{for the 10 most popular tools.} Therein, the results show that \ra{none of these tools are perceived positively in all aspects. Especially, the practitioners reflect more positive opinions on the \textit{architecture} perspective of all these tools; however, more negative on the \textit{measurement} perspective. To be emphasized, such results only indicate, for example, there are more negative opinions on tools' measurement than positive. It does not mean the deficit in these tools' measurement quality.} It is depending on the teams' main interests and their criticality criteria that a selected tool can help them the most with its provided benefits. To be noted, \ra{it is highly likely there are certain amount of irrelevant data being taken into account due to the limitation in the data extraction strategy. The filtering procedure shall help to eliminate the influence yet the outcome can still contain distraction. Nonetheless, including more data sources, such as, forum posts, blogs, and tweets, and more facilitation from human expert shall enrich the opinion pool and further reduce the influence of irrelevant data. }
\end{sloppypar} 


\begin{sloppypar}
\textbf{Which tools are popular?} We observed that the communities behind the analyzed tools differ significantly. In particular, \toolzip~and \tooljae~are the most cited tools in peer-reviewed publications, while \ra{Splunk, Haystack, Sentry, New Relic, Datadog, Zipkin, Jaeger, OpeneTelemetry, Dynatrace and AppDynamics} ~are the most discussed tools in the selected online media (\textbf{RQ$_2$}). Moreover, the community behind Splunk is the most responsive in term of questions answered in the online media channels investing in supporting the community by creating posts and tags. \ra{The other nine popular tools also have reasonably active supporting communities responding to technical issues. To be noted, there is a clear gap in terms of community discussion between the top 10 popular tools and the other tools. However, }
it certainly does not reflect their quality or usefulness to the customers. The difference in popularity may result from marketing strategies, investment in promotion, personal preferences, or simply the psychology of conformity. 
\end{sloppypar}

\ra{To be noted, the different expectation levels of the tool users together with their various requirements can lead to the inevitable difference in their opinions. As the expectation of information system users is ``a set of beliefs held by the targeted users of an information system associated with the eventual performance and with their performance using the system \cite{szajna1993effects}". The initial expectation is formed towards a particular product or service (in our case, tracing tools) by the practitioners. After using the tool, they form the perception of the quality. When such perceived quality is assessed according to their expectation, the extent to which the expectation is confirmed is determined \cite{bhattacherjee2001understanding}. The satisfaction is formed based on the expectation and the corresponding confirmation. In this case, we assume the collective  expectation of the practitioners towards each tracing tool is equal, because 1) the main feature of the tools can be seen as identical and 2) we collected statistically representative amount of data to even the deviation. Therefore, we  assume the collective perceived sentiment of the practitioners towards the tools can be directly compared regardless of the expectation factor. }

\textbf{What is still missing?} It is difficult to identify the benefits and issues in details regarding each identified aspect using the topic modeling and sentiment analysis method. Surveys and interviews of the practitioners or domain expects shall facilitate the further investigation. On the other hand, it is also worthwhile to investigate the in-depth reasons for the high popularity of some tools, e.g., \tooldog~and \toolapp. Furthermore, from the perspective of software evolution, it is also interesting to investigate the changes of these tools over time in terms of the provided features, user perceived benefits and issues, and their popularity.

\section{Threats to Validity}
\label{sec:TV}


Our paper might suffer from threats related to the inaccuracy of the data extraction, a possible incomplete set of results due to limitation of the search terms, bibliographic sources and gray literature search engine, and possible subjectivity related to the definition and the application of the  exclusion/inclusion criteria. 
In the this section, we discuss these threats and the strategies we adopted to mitigate them, based on the standard checklist for validity threats proposed in~\cite{WohlinExperimentation}.

\begin{sloppypar}
\textbf{Construct validity}.
Construct validities are concerned with issues that to what extent the object of study truly represents theory behind the study~\cite{WohlinExperimentation}. The RQs and the classification schema adopted might might suffer of this threat. To limit this threat, the authors reviewed independently and then discussed collaboratively RQs and the related classification schema.
\end{sloppypar}
\noindent To be noted, such the NLP-based approach of detecting the benefits and issues of open tracing tools (\textbf{RQ$_3$}- \textbf{RQ$_4$}) may fall shorts due to the potential data abundance. For some particular tools, e.g., \tooloce, \toolsky~and \toolsta, the related text data is not sufficient compared to that of the other tools, which shall likely result in the lack of reliability in the related conclusion. On the other hand, the selected data sources, such as, Dzone and Medium, aim to introduce and promote emerging or prevailing technologies rather than to criticize them. As a result, the number of texts with negative sentiment is far lower than positive or neutral ones. It is recommended to include more data sources that covers the opinions from the forum end users who provide more unbiased comments and feedback. Furthermore, due to the limitation of LDA topic modeling on short texts, the performance of the approach can be further enhanced by adopting other techniques, such as the Biterm topic model \cite{cheng2014btm}, which shall be included in the future work. 

\textbf{Internal Validity}. The source selection approach adopted in this work is described in Section~\ref{sec:ML}. In order enable the replicability of our work, we carefully identified and reported bibliographic sources adopted to identify the peer-review literature, search engines, adopted for the gray literature,  search strings as well as inclusion and exclusion criteria.
Possible issues in the selection process are related to the selection of search terms that could have lead to a non complete set of results. 
To mitigate this risk, we applied a broad search string. This was possible because of the novelty of the topic.
To overcome the limitation of the search engines, we queried the academic literature from eight bibliographic sources, while we included the gray literature from Google, Medium Search, Twitter Search and Reddit Search. Additionally, we applied a snowballing process to include all the possible sources.
The application of inclusion and exclusion can be affected by researchers’ opinion  and experience. To mitigate this threat, all the sources were evaluated by at least two authors independently. 

\begin{sloppypar}
\textbf{Conclusion validity}.
Conclusion validity is related to the reliability of the conclusions drawn from the results~\cite{WohlinExperimentation}.
To ensure the reliability of our treatments, the terminology adopted in the schema has been reviewed by the authors to avoid ambiguities. All primary sources were reviewed by at least two authors to mitigate bias in data extraction and each disagreement was resolved by consensus, involving the third author.
\end{sloppypar}

\textbf{External Validity.}
External validity is related to the generalizability of the  results of our multivocal literature review.  In this study, we map the literature on Open Tracing Tools, considering both the academic and the gray literature. However, we cannot claim to have screened all the possible literature, since some documents might have not been properly indexed, or possibly copyrighted or, even not freely available.

\section{Related Work}
\label{sec:RW}

\begin{sloppypar}
Among the literature, we identified only one study that---as a primary research goal---compares tools to understand their suitability in different contexts: Li et al. \cite{li2022enjoy} conduct an industrial survey regarding the different adoption strategies of distributed tracing tools. Covering ten different tools and ten different companies, the study finds that the companies' tracing and analysis pipelines are similar and that companies choose different tools based on different concerns and focuses caused by their company size.
\end{sloppypar}

\begin{sloppypar}
More often, researchers compare tracing tools in their state-of-the-art section to point out a research gap and then propose their own approach. For example, Bento et al. \cite{bento2021automated} propose using tracing data to extract service metrics, dependency graphs and work-flows with the objective of detecting anomalous services and operation patterns. Therein, the authors reflect on advantages and disadvantages of the tools \tooljae~and \toolzip~and point out their lack of automated analysis and processing functionality. Along the same lines, Song et al. \cite{song2019astracer} propose ASTracer, a tracing tool for the Apache Hadoop\footnote{\url{https://hadoop.apache.org}} distributed file system. The authors point out the shortcomings of tracing tools like \toolzip, \tooljae~and Htrace\footnote{\url{http://htrace.org}}, e.g., not considering the execution of different call trees or not being able to adapt its sampling rate during execution. 
\end{sloppypar}

\begin{sloppypar}
Some researchers discuss the use of a varying sampling rate when collecting data so that a tool can collect more data when a problem arises and fewer data otherwise. For example, Berg et al. \cite{berg2021snicket} propose Snicket, a distributed tracing system, in which database-style queries are used to express the analysis the developer wants to perform on the trace data. This query is then used to generate microservice extensions that intercept the needed data. The authors compare tracing tools such as Dapper \cite{Dapper2010}, \tooljae, and Canopy \cite{Kaldor2017}, and point out that they may miss important unusual trace information with a uniform, up-front decided sampling rate. They also mention \toollig, which prioritizes latest unusual traces with dynamic tracing and sampling. Also Las-Casas et al. \cite{las2019sifter} propose Sifter, a general-purpose distributed tracing framework that is able to adapt the sampling rate in case of anomalous and outlier executions. Sifter integrates with X-Trace \cite{Fonseca2007}, \tooljae~ and Zipkin to obtain tracing samples. 
\end{sloppypar}

\begin{sloppypar}
Another frequent type of paper is when researchers use tracing tools to obtain traces, which are then used in their research. For example, Gorige et al. \cite{gorige2020privacy} propose a privacy risk detection framework based on distributed tracing, to identify privacy and security risks in microservices. They use \tooljae~ to collect the required data. Iurman et al. \cite{iurman2021towards} propose a unified solution combining in-band telemetry (an approach to collect data about the network state without affecting network performance) and Application Performance Management. In their solution, \tooljae~ and other tracing tools can be used to collect application level information. Avritzer et al. \cite{Avritzer202162} propose PPTAM, a set of tools for performance testing and perfor\-mance-based application monitoring. They also extend their approach in \cite{Avritzer202161} to detect performance anti-patterns.
\end{sloppypar}

\begin{sloppypar}
In summary, within the identified literature, we observe that tracing tools are mentioned to a) develop innovative tracing approaches or b) to enhance existing tools, e.g., allowing an adaptive sampling rate. Moreover, often c) researchers use the collected tracing data to obtain other research goals, e.g., identifying anti-patterns. Rarely, in fact we identified only one, a general-purpose comparison of various tools available on the market is the research goal. Such information can be found often in gray literature, such as blogs or online articles, e.g., \cite{Comparison}, where Barker compares Zipkin, \tooljae, and Appdash\footnote{\url{https://github.com/sourcegraph/appdash}}. Therefore, to obtain an overview over the available tools, a study has to 1) systematically collect the tools discussed in the literature and 2) consider both white and gray literature. This research gap is addressed in this paper. 
\end{sloppypar}

\section{Conclusion}
\label{sec:Conclusion}
In this work, we compared \ra{30} Open Tracing Tools identified by adopting the Systematic Multivocal Literature Review process. For each tool, we investigated the measured features, the popularity both in peer-reviewed literature and online media, and derived benefits and issues. Specially, we adopted topic modeling and sentiment analysis for topic extraction and analysis with ChatGPT to support effective topic interpretation.

The achieved results provided interesting insights among the eleven tools investigated in this study. The deepest comparison we conducted did not allow us to clearly identify a ``silver bullet'' tool for any usage. Each tool has different implications under different conditions. Moreover, it is difficult to identify the benefits and issues in detail regarding each identified aspect using the topic modeling and sentiment analysis method.

\bibliographystyle{model1-num-names}
\bibliography{references.bib,references-gray.bib}

\subsection{Acknowledgment}
The research presented in this article has been partially funded by: the ``Software Rejuvenation'' project funded by Ulla Tuominen Foundation. 


\end{document}